\documentclass[twocolumn,floatfix]{revtex4-2}
\pdfoutput=1
\usepackage{graphicx}
\usepackage{dcolumn}
\usepackage{longtable}
\usepackage{amsmath}
\usepackage{amssymb}

\usepackage{bm}

\begin{document}
\title{The proxy-SU(3) symmetry in atomic nuclei }

\author
{Dennis Bonatsos$^1$, Andriana Martinou$^1$, S.K. Peroulis$^1$, T.J. Mertzimekis$^2$, and N. Minkov$^3$ }

\affiliation
{$^1$Institute of Nuclear and Particle Physics, National Centre for Scientific Research ``Demokritos'', GR-15310 Aghia Paraskevi, Attiki, Greece}

\affiliation
{$^2$  Department of Physics, National and Kapodistrian University of Athens, Zografou Campus, GR-15784 Athens, Greece}

\affiliation
{$^3$Institute of Nuclear Research and Nuclear Energy, Bulgarian Academy of Sciences, 72 Tzarigrad Road, 1784 Sofia, Bulgaria}

\begin{abstract}

The microscopic origins and the up-to-now predictions of the proxy-SU(3) symmetry model of atomic nuclei are reviewed.  Starting from the experimental evidence for the special role played by nucleon pairs with maximal spatial overlap, the proxy-SU(3) approximation scheme is introduced and its validity is demonstrated  through Nilsson model calculations, as well as through its connection to the spherical shell model. The major role played by highest weight irreducible representations of SU(3) in shaping up the nuclear properties is pointed out, resulting in parameter-free predictions of the collective variables $\beta$ and $\gamma$ for even-even nuclei, in the explanation of the dominance of prolate over oblate shapes in the ground states of even-even nuclei, in the prediction of a shape/phase transition from prolate to oblate shapes below closed shells, as well as in the prediction of specific islands on the nuclear chart in which shape coexistence is confined.  Further developments within the proxy-SU(3) scheme  are outlined.

 \end{abstract}

\maketitle
\section{Introduction}  \label{Sec1} 

The purpose of the present article is to discuss the  ideas which led to the introduction of the proxy-SU(3) symmetry model, its microscopic justification, and its applications to nuclear structure problems developed so far. 
Starting with a review of nuclear structure models based on the SU(3) symmetry in Section II, we discuss nucleon pairs favoring deformation in Section III and show how these lead to the introduction of the proxy-SU(3) symmetry in Section IV. The approximation leading to the proxy-SU(3) symmetry is tested using Nilsson model calculations in Section V and is ``translated'' into spherical shell model language in Section VI. After pointing out in Section VII the crucial role played by the highest weight irreducible representations, their consequences on providing parameter-free predictions for the collective variables $\beta$ and $\gamma$ of even-even nuclei, on explaining the dominance of prolate over oblate shapes in the ground states of even-even nuclei and on the prediction of a prolate to oblate shape/phase transition are examined in Section VIII, while in Section IX a mechanism predicting on the nuclear chart specific islands on which shape coexistence can appear is discussed. Possible further work on parameter-free predictions of B(E2) transition rates, energy spectra, binding energies and nucleon separation energies is outlined in Section X.  Literature has been searched up to October 2022.

\section{SU(3) symmetry in nuclear structure} \label{Sec2}

Symmetries have played a major role in nuclear structure since 1937, when Wigner \cite{Wigner1937} suggested that to  a first approximation nuclear forces should be independent of the orientation of the spins and isospins of the nucleons constituting a nucleus, in the framework of what became known as the SU(4) supermultiplet model \cite{Franzini,Pang1969}. 

In 1949 a major step forward in deciphering the experimental observations has been made by Mayer and Jensen through the introduction of the shell model \cite{Mayer1,Mayer2,Jensen,MJ}, who interpreted the appearance of nuclear magic numbers in terms of a three-dimensional (3D) isotropic harmonic oscillator (HO) potential, to which  a crucial spin-orbit term has been added. The shells of the 3D isotropic HO are labeled by the number of excitation quanta $n$, and are characterized by the unitary symmetries U((n+1)(n+2)/2), having SU(3) subalgebras \cite{Wybourne,Smirnov,IacLie,BK}. Mayer and Jensen shared with Wigner the Nobel Prize in Physics in 1963 \cite{Nobel1963}. 

The shell model suffices for describing near-spherical nuclei in the vicinity of the magic numbers, away of which, however, large quadrupole moments are observed. In order to explain their appearance, 
Rainwater in 1950 suggested \cite{Rainwater} that deformed shapes are energetically favored away from closed shells. Soon thereafter in 1952 the collective model of Bohr and Mottelson has been introduced \cite{Bohr,BM}, in which the collective variables $\beta$ and $\gamma$ describe the departure from the spherical shape and from axial symmetry, respectively. Bohr, Mottelson, and Rainwater shared the Nobel Prize in Physics in 1975 \cite{Nobel1975}. 

In 1955 Nilsson introduced \cite{Nilsson1,RN,NR} a modified version of the shell model, using a 3D anisotropic HO with cylindrical symmetry \cite{Takahashi,Asherova,RD,ND,PVI,Lenis}, 
in which axial deformed nuclei with prolate (rugby ball like) or oblate (pancake like) shapes can be described.  

In 1958 Elliott proved that within the nuclear $sd$ shell,  possessing the U(6) symmetry,  deformation can be described in terms of the SU(3) subalgebra \cite{Elliott1,Elliott2,Elliott3,Wilsdon,Elliott4,Harvey,Cseh194}, thus building a bridge between the spherical shell model and the collective model. However, this bridge is valid only in the case of light nuclei, in which the consequences of the spin-orbit interaction on the ordering of the single particle nucleon levels are small. 
Beyond the $sd$ nuclear shell the spin-orbit force \cite{Mayer1,Mayer2,Jensen,MJ} is known to push within each HO shell the orbitals possessing the highest angular momentum $j$ to the shell below. The orbitals left back in each shell after this removal, called the normal parity orbitals, along with the orbitals invading from the shell above, having the opposite parity and called the intruder orbitals, 
 form a new shell, in which  the SU(3) symmetry of the 3D isotropic HO is broken. 

Efforts of extending the SU(3) symmetry to heavy nuclei \cite{Raychev25,Afanasev,Raychev16,RR27}, started in 1972, evolved gradually into the introduction of the Vector Boson Model \cite{Minkov1,Minkov2,Minkov3}. At the same
it has been realized \cite{AR} that the Bohr--Mottelson model has an overall U(5) symmetry possessing an O(5) subalgebra. 

A major step forward in the extention of the SU(3) symmetry to heavy nuclei has been taken in 1973, with the introduction of the pseudo-SU(3) symmetry \cite{Adler,Shimizu,pseudo1,pseudo2,DW1,DW2,Harwood}.
Using a unitary transformation \cite{AnnArbor,Quesne,Hess}, the incomplete set of normal parity orbitals left in a shell is mapped onto the complete set of orbitals of the shell below. In this way the SU(3) symmetry of the 3D isotropic HO is recovered for the normal parity orbitals. This is achieved by assigning to each orbital a pseudo-orbital angular momentum and a pseudo-spin, while keeping the total angular momentum intact. As a result, each shell is formed now by a  normal parity part, which possesses a U(n) symmetry and a SU(3) subalgebra, and by an intruder part which does not possess any SU(3) structure and therefore has to be taken into account separately, using shell model techniques \cite{DW1,DW2}. Later it has been understood \cite{Ginocchio1,Ginocchio2} that the  pseudospin symmetry has relativistic mean field roots.

In 1974 it was realized that the nuclear quadrupole degree of freedom can be described in terms of a SU(6) algebra \cite{Jolos} formed by five generalized coordinates and conjugated momenta, and their commutators.  

Next year Arima and Iachello introduced the Interacting Boson Model (IBM) \cite{AI,AIU5,AISU3,AIO6,IA,IVI,FVI}, in which,  in addition to the quadrupole degree of freedom,  described by $d$-bosons of angular momentum two, the monopole degree of freedom, described in terms of  $s$-bosons of zero angular momentum, is also taken into account. IBM is characterized by an overall U(6) symmetry, possessing three limiting symmetries, U(5) \cite{AIU5} for vibrational nuclei, which corresponds to the Bohr-Mottelson collective model, O(6) \cite{AIO6} for $\gamma$-unstable nuclei (which are soft towards triaxial deformation), and SU(3) \cite{AISU3} for deformed nuclei. 

In 1977 the symplectic model \cite{Rosensteel1,Rosensteel2,Park1984,Rowe1985,RW} was introduced by Rowe and Rostensteel. The symplectic model, having an overall symmetry called Sp(3,R) by Rowe and Rosensteel but Sp(6,R) by other authors \cite{Wybourne1992,Escher1998}, is a generalization of the fermionic Elliott SU(3) model to include many major oscillator shells in addition to core excitations. Its overall symmetry is characterized by the noncompact algebra Sp(6,R), which does possess a compact SU(3) subalgebra along with other ones. A proton-neutron extension of the model, called the proton-neutron symplectic model (PNSM) is also developed \cite{Ganev1,Ganev2,Ganev3}.

In 1982 the Interacting two-Vector Boson Model \cite{IVBM1,IVBM2} has been introduced, described in terms of two vector bosons of angular momentum one.  These form the noncompact algebra Sp(12,R), of which the maximal compact subalgebra is U(6).  

In 1987 the Fermion Dynamical Symmetry Model \cite{FDSM} has been introduced. Instead of using the usual splitting of the total angular momentum of the nucleons into orbital angular momentum and spin parts, in this model a splitting of  the total angular momentum into active and inactive parts is assumed, the k-active part containing a SU(3) subalgebra. 

In 2000 an {\it ab initio} approach to a no-core shell model has been introduced \cite{nocore1,nocore2} for light nuclei. It has been soon thereafter realized that a symplectic symmetry is underlying the {\it ab initio} no core shell model results, thus paving the way towards the development of the symplectic no-core shell model \cite{nocore3,nocore4,DytrychJPG2008}, which has soon been extended to intermediate-mass nuclei \cite{Tobin2014,DytrychCPC2016,LauneyIJMPE2015,LauneyPPNP2016}. The realization that nuclei are made of only a few equilibrium shapes \cite{DytrychPRL2020} led to the introduction of the {\it ab initio} symmetry-adapted no-core shell model \cite{LauneyEPJST2020,LauneyARN2021}. A SU(3)-adapted basis \cite{DytrychCPC2016} plays a key role in this approach, taking advantage of the Elliott SU(3) symmetry underpining the Sp(3,R)                              
   \cite{Rosensteel1,Rosensteel2,Park1984,Rowe1985,RW} (alias Sp(6,R) \cite{Wybourne1992,Escher1998}) symplectic model.  

The use of the  SU(3) symmetry in nuclear structure has been reviewed in 2020 by Kota \cite{Kota}. 

In 2017 the proxy-SU(3) symmetry has been suggested \cite{proxy1,proxy2,proxy3}, which will be the subject of the present review. However, before describing the proxy-SU(3) symmetry itself, we briefly review the physical motivation behind its introduction. 

\section{Nucleon pairs favoring deformation}  \label{Sec3} 

In 1953 deShalit and Goldhaber \cite{deShalit} in their studies of $\beta$ transition probabilities noticed that the development of nuclear deformation was favored within the proton--neutron pairs of orbitals (1p3/2, 1d5/2), (1d5/2, 1f7/2), (1f7/2,   1g9/2), (1g9/2, 1h11/2), (1h11/2, 1i13/2), since the nucleons of one kind (protons, for example) appeared to have a stabilizing effect on pairs of nucleons of the other kind (neutrons in the example). In the standard shell model notation $\vert n l j m_j\rangle$, states are labeled by the number of oscillator quanta $n$, the orbital angular momentum $l$, the total angular momentum $j$, and its $z$-projection $m_j$. In this notation these orbitals form pairs differing by   $\vert \Delta n \Delta l \Delta j \Delta m_j\rangle = \vert 0 1 1 0\rangle$, which  we are going to call the spherical shell model $\vert 0 1 1 0\rangle$ pairs, or simply the $\vert 0 1 1 0\rangle$ pairs. 

In 1962 Talmi introduced the concept of seniority  \cite{Talmi62,Talmi71,Talmi73,Talmi93}, corresponding to the number of nucleon pairs coupled to non-zero angular momentum. Seniority explained the linear dependence of neutron separation energies on the mass number within various series of isotopes, being a major step forward in our understanding of effective interactions and coupling schemes in nuclei. 

In 1977 Federman and Pittel \cite{FP1,FP2,FP3} realized that certain proton--neutron pairs play a major role when adding valence protons and valence neutrons to a nucleus. In particular, the proton--neutron pairs (1d5/2, 1d3/2), (1g9/2, 1g7/2), (1h11/2, 1h9/2), and (1i13/2, 1i11/2) are important for the onset of deformation, while further on deformation is reinforced by the proton--neutron pairs (1d5/2, 1f7/2), (1g9/2, 1h11/2), (1h11/2, 1i13/2), and (1i13/2, 1j15/2). In the shell model notation these sets of pairs, which are shown in Table I, correspond to $\vert \Delta n \Delta l \Delta j \Delta m_j\rangle = \vert 0 0 1 0\rangle$ and $\vert 0 1 1 0\rangle$ respectively. The latter set coincides with the de Shalit--Goldhaber pairs mentioned above. 

In 1985 Casten demonstrated the decisive role played by proton-neutron pairs through the introduction of the $N_p N_n$ scheme \cite{CastenPRL,CastenNPA}. This scheme showed the systematic dependence of several observables on the quadrupole-quadrupople interaction, which is measured through the product $N_pN_n$, where $N_p$ ($N_n$) is the number of valence protons (neutrons) counted from the nearest closed shell. In 1987 the systematic dependence of several observables on the $P$-factor, defined as $P= N_p N_n / (N_p+N_n)$ \cite{Haustein,Castenbook}, has been demonstrated. The $P$-factor expresses the competition between the quadrupole deformation and the pairing interaction, ``measured'' by $N_pN_n$ and  $N_p+N_n$, respectively.  

In 1995 the proton--neutron pairs (1g9/2, 2d5/2), (1h11/2, 2f7/2), (1i13/2, 2g9/2) led to the introduction of the quasi-SU(3) symmetry \cite{Zuker1,Zuker2}, which leads to enhanced quadrupole collectivity \cite{Kaneko}. 
 In the shell model notation these pairs are expressed as $\vert \Delta n \Delta l \Delta j \Delta m_j\rangle = \vert 1 2 2 0\rangle$.

Detailed studies of double differences of experimental binding energies \cite{Cakirli94,Cakirli96,Brenner,Cakirli102}, led in 2010 to understanding \cite{Burcu} that proton-neutron pairs differing by $\Delta K[ \Delta N \Delta n_z \Delta \Lambda]=0[110]$ in the Nilsson notation \cite{Nilsson1,RN,NR} $K [N n_z \Lambda]$, where $N$ is  the total number of oscillator quanta, $n_z$ is the number of quanta along the $z$-axis, and $\Lambda$, $K$ are the projections along the $z$-axis of the orbital angular momentum and the total angular momentum respectively, play a major role in the development of nuclear deformation. This effect has been attributed to their large spatial overlaps \cite{Karampagia} and has been corroborated by nuclear density functional theory calculations \cite{Stoitsov}. We are going to call these pairs the Nilsson 0[110] pairs, or simply the 0[110] pairs. Notice the difference in the notation in comparison to the spherical shell model $\vert 0 1 1 0\rangle$ pairs. 

As we shall see below, the Nilsson 0[110] pairs play a crucial role in the replacements made within the approximation leading to the proxy-SU(3) symmetry. Furthermore, the relation between the Nilsson 0[110] pairs and the spherical shell model $\vert 0 1 1 0\rangle$ pairs will be clarified through the connection of the proxy-SU(3) symmetry to the spherical shell model. Evidence supporting the formation of 0[110] pairs has been found recently within Monte Carlo shell model calculations \cite{Sieja}. 

\section{The proxy-SU(3) approximation}  \label{Sec4}

The proxy-SU(3) symmetry has been born when the desire to reestablish the SU(3) symmetry of the 3D-HO in medium mass and heavy nuclei, described in section \ref{Sec2}, met the experimental hint of the 0[110] Nilsson pairs, described in section \ref{Sec3}. The 0[110] Nilsson pairs have been discovered through the experimental observation \cite{Burcu} that proton-neutron pairs of this type maximize the proton-neutron interaction. This maximization has been attributed to the large spatial overlaps of the two orbitals involved in such pairs \cite{Karampagia}. In other words, 0[110] Nilsson pairs are very similar, since they possess identical angular momentum and spin properties (identical projections of the orbital angular momentum, the spin, and the total angular momentum) and very similar spatial shapes, since they differ only by one excitation quantum in the $z$-axis (and therefore also by one quantum in  the total number of quanta). The similarity holds equally well if one considers a 0[110] pair of protons, or a 0[110] pair of neutrons. Such pairs will still have identical angular momentum and spin properties and very similar spatial shapes. Therefore, in the framework of an approximation needed for some reason,  each of the members of a 0[110] pair could replace the other, i.e., act as its proxy, with minimal changes inflicted in the physical system under study. 

Such a situation appears in the shells of the shell model beyond 28 nucleons. The intruder orbitals, which have come down from the shell above, form 0[110] pairs with the orbitals which have deserted the shell by fleeing into the shell below. We can therefore think of replacing the intruder orbitals by their 0[110] counterparts, which are the deserting orbitals, expecting that the changes caused in the physical system under study would be minimal. In other words, the deserting orbitals can act as proxies of the intruder orbitals. But in this way the shell gets back the SU(3) symmetry of the 3D-HO, which it had lost after the deserting of the orbitals which fled to the shell below. 

An example can be seen in Fig. 1, in which the $sdg$ shell is depicted, consisting of the $3s_{1/2}$, $2d_{3/2}$, $2d_{5/2}$, $1g_{7/2}$, $1g_{9/2}$ orbitals, having a U(15) overall symmetry, which possesses a SU(3) subalgebra. The spin-orbit interaction pushes the orbital with the highest $j$, i.e., $1g_{9/2}$, to the shell below, while it brings down the $1h_{11/2}$ orbital from the shell above. In this way the 50-82 shell of the shell model is formed, in which the SU(3) symmetry has been lost. Looking however into the details of the orbitals, a sign of hope appears. The intruder $1h_{11/2}$ orbital consists of the Nilsson orbitals 1/2[550], 3/2[541], 5/2[532], 7/2[523], 9/2[514], 11/2[505],  while the deserting $1g_{9/2}$ orbital consists of the Nilsson orbitals 1/2[440], 3/2[431], 5/2[422], 7/2[413], 9/2[404]. We observe that the first five Nilsson orbitals of $1h_{11/2}$ are 0[110] partners with the five orbitals making up $1g_{9/2}$. Therefore it is plausible to replace the 1/2[550], 3/2[541], 5/2[532], 7/2[523], 9/2[514] orbitals by the  1/2[440], 3/2[431], 5/2[422], 7/2[413], 9/2[404] orbitals, hoping that the changes inflicted in the physical system would be minimal. In other words, the 1/2[440], 3/2[431], 5/2[422], 7/2[413], 9/2[404] orbitals will act as proxies of the 1/2[550], 3/2[541], 5/2[532], 7/2[523], 9/2[514] orbitals. 

It is now time to consider the gains and losses of this replacement. The gain is that the SU(3) symmetry has been reestablished in the 50-82 shell, since the shell will now contain all the orbitals composing the $sdg$ shell. However, a couple of losses are lurking. First, the 11/2[505] Nilsson orbital had no 0[110] partner, thus it is left behind within the 50-82 shell, where it stays outside the SU(3) symmetry and in principle it has to be treated separately by shell model techniques \cite{DW1,DW2}. The good news in this case is that the 11/2[505] Nilsson orbital lies at the very top of the 50-82 shell, as one can see in the standard Nilsson diagrams \cite{NR,Lederer}. Thus it should be empty for most nuclei and therefore have little influence on the structure of most of them. The second problem arises from the fact that the intruder orbitals and the deserting orbitals have opposite parities, since they belong to adjacent shells differing by one unit in $N$. For even-even nuclei this should not cause any trouble, since, for example,  a pair of 9/2[514] particles will be replaced by a pair of 9/2[404] particles, therefore no differences caused by parity should be seen. For odd nuclei one would probably have to use projection techniques \cite{Ring}, a problem which has not been considered yet. 

\section{Corroboration of proxy-SU(3) through Nilsson model calculations} \label{Sec5}

The first test \cite{proxy1} of the accuracy of the proxy-SU(3) approximation described in the previous section has been performed in the framework of the Nilsson model. In each shell of the shell model, two calculations have been performed, one with the real orbitals composing the shell, and another one with the intruder orbitals replaced by their 0[110] counterparts. Numerical results for the 82-126 shell are shown in Table II. In the upper part of the table, the matrix elements of the Nilsson Hamiltonian, using the standard set of parameters \cite{BM}, are shown, while in the lower part the matrix elements occurring within the $pfh$ shell resulting after the proxy-SU(3) approximation are reported. We remark that in the last 7 columns of the upper part, corresponding to the $1i_{13/2}$ orbitals, all non-diagonal matrix elements vanish, due to the fact that they connect orbitals of opposite parity. On the contrary, in the last 6 columns of the lower part of the table, in which the proxies of the $1i_{13/2}$ orbitals, i.e., the $1h_{11/2}$ orbitals appear, not all off-diagonal matrix elements vanish, since in this case they connect orbitals of the same parity. However, the number of these off-diagonal matrix elements is small, and in addition their sizes are small in comparison to the diagonal matrix elements, thus they are not expected to affect the single particle energy levels substantially. The diagonal matrix elements in the last six columns of the lower part of the table are also slightly modified in comparison to the corresponding matrix elements in the upper half of the table, but again the changes are small and are not expected to affect the single particle energy levels radically. The evolution of the single particle energy levels with deformation, depicted in Fig. 2, shows that the changes caused in the Nilsson diagrams by the proxy-SU(3) approximation are minimal. Neither the order of the orbitals, nor their dependence on the deformation (upsloping or downsloping) are modified. Similar tables and figures for other shells can be found in the supplemental material for Ref. \cite{proxy1}. It is seen there that the quality of the proxy-SU(3) approximation becomes better in higher shells.  
  
\section{Proxy-SU(3) symmetry in the spherical shell model basis}  \label{Sec6} 

We have seen that the proxy-SU(3) symmetry is established in the nuclear shells beyond the $sd$ shell in the framework of the Nilsson model by taking advantage of the 0[110] Nilsson pairs. The question arises if such a process, allowing the replacement of certain orbitals by their proxies in order to reestablish the SU(3) symmetry, is possible within the framework of the spherical shell model. 

In order to examine if such a possibility exists, we start with the Elliott model \cite{Elliott1,Elliott2,Elliott3,Elliott4,Harvey}, in which  the cartesian basis of the 3D isotropic HO is used. This basis is labeled as $[n_z n_x n_y m_s]$, in which the number of quanta along the $z$, $x$, $y$ directions and the $z$-projection of the spin appear. The first step is to transform this basis into the spherical basis $[n l m_l m_s]$ in $l$-$s$ coupling, labeled by the principal quantum number $n$, the orbital angular momentum $l$, its $z$-projection ($m_l$), and the $z$-projection of the spin ($m_s$). This can be achieved through a unitary transformation \cite{Davies,Chasman,Chacon} 
\begin{equation}
[n_z n_x n_y m_s] = R [n l m_l m_s], 
\end{equation}
the details of which can be found in Ref. \cite{EPJASM}.   Furthermore, the spherical basis can be recoupled from the $l$-$s$ coupling to the $j$-$j$ coupling  through the use of Clebsch-Gordan coefficients \cite{Edmonds,Varshalovich}
\begin{equation}
[n l m_l m_s] =  C [n l j m_j],  
\end{equation}
where the total angular momentum $j$ and its $z$-projection appear. Combining these two transformations one obtains the desired connection between the cartesian Elliott basis and the spherical shell model basis in $j$-$j$ coupling
\begin{equation}
[n_z n_x n_y m_s] = R  C [n l j m_j].  
\end{equation}
 In Ref. \cite{EPJASM} one can find details of the calculations and transformation tables up to the $N=3$  shell. 

The above transformation means that the Nilsson 0[110] replacements made within the proxy-SU(3) scheme correspond to $| 0 1 1 0\rangle$ replacements within the spherical shell model basis. In Table III the correspondence between  original shell model orbitals and proxy-SU(3) orbitals is summarized,  paving the way for using the proxy-SU(3) symmetry in shell model calculations for heavy nuclei, in a way similar to the recently developed for light nuclei  symmetry-adapted no-core shell model approach \cite{LauneyIJMPE2015,LauneyPPNP2016}. 

By analogy to the unitary transformation occurring in the framework of the pseudo-SU(3) symmetry \cite{AnnArbor,Quesne,Hess}, a unitary transformation connecting the orbitals being replaced within the proxy-SU(3) scheme has been found \cite{EPJASM} within the shell model basis and is depicted in Fig. 3.  

The above transformation indicates  that the 0[110] Nilsson pairs identified in Ref. \cite{Burcu} and used within the proxy-SU(3) scheme \cite{proxy1,proxy2,proxy3} are identical to the above mentioned de Shalit--Goldhaber pairs \cite{deShalit} and the Federman--Pittel pairs \cite{FP1,FP2,FP3},  which are expressed as $| 0 1 1 0\rangle$ within the spherical shell model basis. 

Further corroboration of the correspondence between Nilsson pairs and shell model pairs has been provided by calculations \cite{EPJPSM} within the Nilsson model, in which the first justification of the proxy-SU(3) scheme has been found, as mentioned in Sec. \ref{Sec5}. The replacements used in proxy-SU(3) work only for the Nilsson orbitals which possess the highest total angular momentum $j$ within their shell, as one can see in Tables IV and V.

\section{The dominance of the highest weight irreducible representations of SU(3)} \label{Sec7}

In order to start examining the consequences of the existence of the proxy-SU(3) symmetry in medium mass and heavy nuclei, we should first consider a few properties of the irreducible representations (irreps) of SU(3). 

In Elliott's notation \cite{Elliott1,Elliott2,Elliott3,Wilsdon,Elliott4,Harvey} the irreps of SU(3) are labeled by $(\lambda,\mu)$, where the Elliott quantum numbers $\lambda$ and $\mu$ are connected to the number of boxes in the corresponding Young diagram through the relations 
\begin{equation}
\lambda=f_1-f_2, \qquad  \mu=f_2, 
\end{equation}
where $f_1$ ($f_2$) is the number of boxes in the first (second) line of the relevant Young diagram. Irreps with $\lambda > \mu$ are known to correspond to prolate (rugby ball like) shapes, while irreps with $\lambda < \mu$ are known to describe oblate (pancake like) shapes, with $\lambda=\mu$ irreps corresponding to maximally triaxial shapes \cite{Castanos,Evans,Park}. 

A quantity characterizing the SU(3) irreps is the eigenvalue of the second order Casimir operator of SU(3), given by  \cite{Wybourne,IacLie}
\begin{equation}\label{Casimir}
C_2(\lambda,\mu) ={2\over 3}  (\lambda^2 + \mu^2 + \lambda \mu + 3\lambda + 3\mu). 
\end{equation}
This quantity is known to be connected to the eigenvalues of the quadrupole-quadrupole interaction by \cite{Elliott1,Elliott2,Elliott3,Harwood}
\begin{equation} \label{QQC2}
QQ= 4 C_2 -3 L(L+1), 
\end{equation} 
where $L$ is the eigenvalue of the orbital angular momentum. For the ground state bands of even-even nuclei, in which $L=0$, it is clear that the eigenvalue of $C_2$ is proportional to the eigenvalue of the quadrupole-quadrupole interaction, $QQ = 4 C_2$. Since the quadrupole-quadrupole interaction is known to cause nuclear deformation, it is expected that the irrep with the highest value of $C_2$ should correspond to the preferred ground state with maximum deformation. 

There is, however, another factor which should be taken into account. The nucleon-nucleon interaction is known to be attractive and of short range \cite{Mayer1950}, therefore favoring the maximal spatial overlap \cite{Karampagia} among the nucleons, which can be achieved for the most symmetric SU(3) irrep. In the Young diagrams it is known that boxes on the same line correspond to symmetrized particles, while boxes in the same column correspond to antisymmetrized particles \cite{Wybourne,IacLie}. The degree of symmetrization of a given SU(3) irrep can therefore be measured by the ratio of the symmetrized boxes over the total number of boxes, which is 
\begin{equation}
r= {f_1 \over f_1+f_2}= {\lambda +\mu \over \lambda +2\mu}. 
\end{equation}
It has been proved \cite{EPJAHW} that the irreps with the highest value of $r$, i.e., with the highest degree of symmetry, correspond to what is called in the mathematical language the highest weight (hw) irreps of SU(3). As such these irreps appear favored by the nucleon-nucleon interaction and, therefore, dominate the related nuclear properties.

The irreps possessing the highest eigenvalue of the second order Casimir operator, to be called in what follows the highest $C_2$ irreps for brevity, and the highest hw irreps of SU(3) in the various nuclear shells are shown in Table VI.  It is clear that up to midshell the $C_2$ and hw irreps are identical, while in the upper half of each shell the $C_2$ and hw irreps are different, with the exception of the last 5 particle numbers (which correspond to states with up to 4 holes at the top of the shell). Further mathematical details on the dominance of the hw irreps can be found in Refs. \cite{40GC,78IC}. 

\section{Physical consequences of the dominance of the highest weight irreps}  \label{Sec8}

\subsection{Prolate over oblate dominance}

One of the long standing puzzles in nuclear structure is the dominance of prolate (rugby ball like) shapes over oblate (pancake like) shapes in the ground state bands of even-even nuclei. In Elliott's notation, prolate (oblate) irreps correspond to $\lambda>\mu$ ($\lambda<\mu$) \cite{Elliott1,Elliott2,Elliott3,Wilsdon,Elliott4,Harvey,Guzman}, while irreps with $\lambda=\mu$ correspond to maximal triaxiality.  Despite several attempts \cite{Castenbook,Hamamoto,Tajima64,Tajima702,Tajima86} to resolve this puzzle, the question is still considered open \cite{HM}.

The dominance of the hw irreps over the highest $C_2$ irreps offers a simple, parameter-free justification of the prolate over oblate dominance. As an example, in Table VII, the hw irreps corresponding to the rare earth nuclei with valence protons in the 82-126 shell and valence neutrons in the 126-184 shell are shown. These hw irreps are obtained  by adding up the hw irrep $(\lambda_\pi, \mu_\pi)$ corresponding to the valence protons and the hw irrep $(\lambda_\nu, \mu_\nu)$ corresponding to the valence neutrons into the most stretched irrep \cite{DW1} $(\lambda_\pi+\lambda_\nu, \mu_\pi+\mu_\nu)$. We notice that prolate irreps are obtained over most of the table, with the exception of its lower right corner, near which a few oblate irreps (underlined in Table VII) appear in nuclei lying below the top of both the proton valence shell and the neutron valence shell.  The same effect appears in other shells as well, as one can see using the irreps appearing in Table VI. For example, results for rare earths with valence protons in the 50-82 shell and valence neutrons within the 50-82 and the 82-126 shells are given in Ref. \cite{proxy2}. The conclusion of this subsection has been recently corroborated by a heuristic method in Ref. \cite{Sugawara106}. 
 
\subsection{Parameter-free predictions for the collective variables $\beta$ and $\gamma$} \label{subsec82}

Further consequences of the dominance of the hw irreps over the highest $C_2$ irreps become evident if one considers the connection between the Elliott quantum numbers $\lambda$, $\mu$ and the collective variables $\beta$, $\gamma$ of the Bohr--Mottelson model. This connection is obtained by employing a linear mapping between the invariant quantities of the two theories, which are the invariants $\beta^2$ and $\beta^3 \cos 3 \gamma$ of the collective model \cite{Bohr,BM} and the Casimir operators of second and third order of SU(3) \cite{Wybourne,IacLie}. 
This mapping provides for the angle collective variable $\gamma$ the expression \cite{Castanos,Park}
\begin{equation}\label{g1}
\gamma = \arctan \left( {\sqrt{3} (\mu+1) \over 2\lambda+\mu+3}  \right),
\end{equation}
while for the square of the deformation parameter $\beta$ it gives  \cite{Castanos,Park}
\begin{equation}\label{b1}
	\beta^2= {4\pi \over 5} {1\over (A \bar{r^2})^2} (\lambda^2+\lambda \mu + \mu^2+ 3\lambda +3 \mu +3), 
\end{equation}
which indicates that $\beta^2$ is proportional to the second order Casimir operator of SU(3) \cite{IA}. In Eq. (\ref{b1})
$A$ is the mass number of the nucleus, while $\bar{r^2}$ is related to the dimensionless mean square radius \cite{Ring}, $\sqrt{\bar{r^2}}= r_0 A^{1/6}$, which  
is obtained by dividing the mean square radius, being proportional to $A^{1/3}$, by the oscillator length, which grows as $A^{1/6}$ \cite{Ring}.
For the constant $r_0$ the value 0.87 is used, determined from a fit over a wide range of nuclei \cite{DeVries,Stone}, 
as done in Ref.  \cite{Castanos}.   

A word of warning is in place here. Since in proxy-SU(3) only the valence nucleons are taken into account, the values of $\beta$ obtained from Eq. (\ref{b1}) should be rescaled by multiplying them by a factor $A/(S_\pi+S_\nu)$, where $S_\pi$ ($S_\nu$) is the size of the proton (neutron) valence shell. The need for this rescaling has been discussed in detail in Sec. V of Ref. \cite{proxy2}. In practice this rescaling means that  Eq. (\ref{b1}), when used in the proxy-SU(3) framework, should read 
\begin{equation}\label{b2}
	\beta^2= {4\pi \over 5} {1\over ((S_\pi+S_\nu) \bar{r^2})^2} (\lambda^2+\lambda \mu + \mu^2+ 3\lambda +3 \mu +3). 
\end{equation}

Numerical results for the collective variable $\beta$ ($\gamma$) for several rare earth nuclei are shown in Fig. 4 (Fig. 5) and are compared to other theoretical approaches, like the D1S-Gogny interaction \cite{Gogny} and relativistic mean field theory \cite{LalazissisADNDT71}, as well as to experimental values \cite{Raman}. By ``Gogny D1S min.'' we label the values of the deformation variables $\beta$ and $\gamma$ at the HFB minimum energy (entries 4 and 5 in Ref. \cite{Gogny}), while by ``Gogny D1S mean'' the mean ground state values (entries 11 and 12 in Ref. \cite{Gogny}). It is seen that the {\sl parameter-free} proxy-SU(3) predictions are in good agreement with both the theoretical approaches and the empirical values. Additional numerical results for $\beta$ and $\gamma$ obtained within the proxy-SU(3) approach can be found in \cite{34J,EC12,41GC}. Additional comparisons of proxy-SU(3) predictions to covariant density functional theory can be found in Refs. \cite{Awwad,Alstaty,Elsharkawy}, while comparisons to recent experimental findings can be found in \cite{Canavan,Knafla}. 

In Fig. 4 it is clear that the $\beta$ curve is not symmetric around midshell, but it appears to exhibit higher values in the first half of the shell. The origin of this discrepancy can be traced in Fig. 6, in which the square root of $C_2$, which is proportional to $\beta$, according to Eq. (\ref{b2}), is shown. We see that the breaking of the symmetry around midshell is due to the fact that in the upper half of the shell the highest weight irreps enter in the place of the highest $C_2$ irreps, as indicated by Table 6. 
 
\subsection{Prolate to oblate shape/phase transition}  
 
A second important consequence of the hw irreps dominance is seen in Fig. 7, in which the proxy-SU(3) predictions for the collective variable $\beta$ for the rare earths with valence protons in the 50-82 shell and valence neutrons in the 82-126 shell are collected. The dip seen at $N=116$ signifies the occurrence of a shape/phase transition \cite{Deans,IacE5,CZE5,IacX5,CZX5,IacIJMPA,Poenaru,McCutchan34,JolieRMP} from prolate to oblate shapes, for which extended experimental evidence exists \cite{Namenson,Alkhomashi,Wheldon,Podolyak,Linnemann}, along with microscopic theoretical considerations \cite{Kumar1,Kumar2,Sarriguren77,Robledo36,Nomura83,Nomura84,Sun659}, and relevant searches within the interacting boson model (IBM) \cite{Jolie1,Jolie2,Thiamova765,Bettermann81,Zhang40,Zhang85} and the Bohr--Mottelson collective model \cite{Z5,Z4}. In the framework of the Bohr Hamiltonian this shape/phase transition has been referred to as Z(5) \cite{Z5}. {\it A posteriori} corroboration of the findings of Ref. \cite{proxy2} is found in Refs. \cite{Alimohammadi,Mutsher,Bindra}.   

The robustness of this result is emphasized by the fact that it also appears in atomic clusters \cite{Clemenger,deHeer,Brack,Nester,deHeer2,Greiner}. The valence electrons in alkali metal clusters, in particular, are supposed to be free, thus forming shells with major magic numbers 2, 8, 20, 40, 58, 92, \dots \cite{Martin1,Martin2,Bjorn1,Bjorn2,Knight1,Peder,Brec1,Brec2}.  Prolate  and oblate shapes in alkali metal clusters have been observed experimentally through optical response measurements, finding oblate shapes below cluster sizes 20 and 40 \cite{Borggreen,Pedersen1,Pedersen2,Haberland}, while prolate shapes have been seen above cluster sizes 8, 20, 40 \cite{Borggreen,Pedersen1,Pedersen2,Haberland,Schmidt}. In other words, prolate (oblate) shapes are seen above (below) the magic numbers, exactly as in atomic nuclei.  

It should be noticed that the shape/phase transition from prolate to oblate shapes is seen equally clearly in the framework of pseudo-SU(3), if the dominance of the hw irreps over the highest $C_2$ irreps is taken into account, as discussed in detail in Ref. \cite{EPJST}. Although pseudo-SU(3) and proxy-SU(3) are based on different approximations, involving different unitary transformations, they lead to the same physical conclusion, 
a fact providing evidence for the compatibility of the two approaches. 

The compatibility of the proxy-SU(3) and pseudo-SU(3) approaches has also been pointed out recently \cite{Cseh101,Hess57,Hess104} in the framework of the semimicroscopic algebraic quartet model (SAQM) \cite{Cseh743}, 
in which quartets consisting of two protons and two neutrons occupying a single orbital are taken into account, their excitation spectra occurring when a quarter jumps as a whole to the next major shell. 
 Based on the SU(3) symmetry, in its initial form \cite{Cseh743} the model had been applicable only to light nuclei, based on the Elliott SU(3) symmetry. However, it has recently \cite{Cseh101,Hess57,Hess104} been extended to heavy nuclei, based on the proxy-SU(3) symmetry and on the pseudo-SU(3) symmetry, with similar results provided by both approaches. The use of proxy-SU(3) could also be extended in a similar way to the  Semimicroscopic Algebraic Cluster Model (SACM) \cite{Cseh281,Levai230,Hess1016}, in which the internal structure of the clusters is described in terms of the shell model SU(3) symmetry, while their relative motion is described in terms of the phenomenological algebraic vibron model \cite{FVI,Levine77,vanRoos79,Daley967}.

\section{Islands of shape coexistence} \label{Sec9}

\subsection{Harmonic oscillator (HO) and spin-orbit (SO) magic numbers}

Shape coexistence (SC) is called the appearance in a nucleus of two bands lying close in energy but having radically different structures, for example one of them being spherical and the other deformed, or both of them being spherical, but one of them exhibiting prolate (rugby ball like) deformation and the other one oblate (pancake like) deformation. Shape coexistence has first been suggested in 1956 by Morinaga \cite{Morinaga}, in relation to the spectrum of $^{16}$O. Since then many experimental examples have been found in both odd and even  nuclei, as summarized in the relevant review articles \cite{Meyer102,Wood215,HW,Garrett124}. From the theoretical point of view, SC has been attributed to the existence of particle-hole excitations across shell or subshell closures, and has been believed to be able to appear all over the nuclear chart, although in Fig. 8 of the authoritative review article by Heyde and Wood \cite{HW} the regions in which SC has been experimentally observed appear to form certain islands on the nuclear chart.

As it has already been mentioned, SC is supposed to be due to particle-hole excitations across shell or subshell closures. However, the magic numbers of the shell model, 2, 8, 20, 28, 50, 82, 126, \dots, are known to be valid only at zero deformation. These magic numbers originate from the 3D-HO magic numbers 2, 8, 20, 40, 70, 112, 168, \dots, \cite{Wybourne,IacLie} (to be called the HO magic numbers in what follows) because of the action of the spin-orbit interaction \cite{Mayer1,Mayer2,Jensen,MJ}. As deformation sets in, the energy gaps separating different shells soon disappear, as one can see in the standard Nilsson diagrams \cite{Nilsson1,RN,NR,Lederer}
(see also Refs. \cite{Sorlin,Otsuka92} for the evolution of magic numbers away from stability). Furthermore, as seen in Table III (see also Ref. \cite{8V}), in the framework of the proxy-SU(3) symmetry a set of magic numbers 6, 14, 28, 50, 82, 126, 184, \dots (to be called the SO magic numbers in what follows) appears, corresponding to the strong presence of the spin orbit interaction everywhere. We remark that the standard shell model magic numbers follow the HO magic numbers up to 20, while they follow the SO magic numbers beyond this point.  

\subsection{A dual shell mechanism for shape coexistence}

It has been suggested \cite{44GC,EPJASC,HINP2021} that the interplay of HO and SO magic numbers offers a simple justification for the appearance of islands of SC on the nuclear chart. Let us see how this is occurring. 

Because of the collapse of the shell model quantum numbers as deformation sets in, the protons and/or the neutrons of a nucleus can follow either the HO or the SO magic numbers. The same number of protons or neutrons will then correspond to a different irrep in the HO framework and to another irrep in the SO framework. These irreps can be seen in Table VIII. As an example, let us consider 60 nucleons. In the HO framework, 40 is a magic number, thus there are left 20 nucleons in the 40-70 shell, which has the U(15) symmetry, and therefore the 20 nucleons correspond to the (20,0) hw irrep according to Table VI. In the SO framework, 50 is a magic number, thus there are 10 nucleons left in the 50-82 shell, which in the proxy-SU(3) approximation has the U(15) symmetry, therefore the 10 nucleons correspond to the (20,4) hw irrep of SU(3) according to Table VI. Indeed in Table VIII for 60 nucleons the hw irreps given are (20,4) for the SO case and (20,0) for the HO case. 

For the $L=0$ bandheads of two coexisting bands one can use the very simple Hamiltonian \cite{EPJASC}
\begin{equation}
H = H_0 -{\kappa\over 2} QQ,  
\end{equation}  
where $H_0$ corresponds to the 3D-HO Hamiltonian and $QQ$ to the quadrupole-quadrupole interaction. Both bands should belong to the same U(n) algebra within the SO and HO schemes. From Table III we see that this is possible for the nucleon number intervals 6-8, 14-20, 28-40, 50-70, 82-112, 126-168, in which both bands belong to the U(3), U(6), U(10), U(15), U(21), U(28) algebra respectively. We remark that the right borders of these regions are the HO magic numbers. 
The successful parameter-free predictions of the $\beta$ and $\gamma$ collective variables for the ground states of nuclei seen in Sec. \ref{subsec82} imply that the ground-state band will belong to the SO irrep, thus the bandhead of the coexisting band should lie higher in energy.  One can easily see that this requirement leads to the condition \begin{equation}
QQ_{SO} \geq QQ_{HO}, 
\end{equation}
the full details of the argument given explicitly in Sec. 8 of Ref. \cite{EPJASC}.  From Eq. (\ref{QQC2}) one then sees that this condition is equivalent to the condition 
\begin{equation}\label{ineq}
C_2(\lambda_{SO},\mu_{SO}) \geq C_2(\lambda_{HO},\mu_{HO}).  
\end{equation}
The eigenvalues of the Casimir operator $C_2$ in the SO and HO frameworks are shown for the various shells in Fig. 8. We see that the condition of Eq. (\ref{ineq}) starts being fulfilled at the nucleon numbers 7, 17, 34, 59, 96, 146, which will therefore stand for the left borders of the regions in which SC could be possible, the right borders being given by the HO magic numbers mentioned above. Therefore we conclude that SC can occur only within the nucleon intervals 7-8, 17-20, 34-40, 59-70, 96-112, 146-168, bearing the U(3), U(6), U(10), U(15), U(21), U(28) symmetry respectively. These intervals define horizontal and vertical stripes on the nuclear chart, shown in color in Fig. 9. One can easily see that the islands of SC seen in Fig. 8 of Ref. \cite{HW} do lie within the stripes predicted by the dual shell mechanism within the proxy-SU(3) framework just found. 

\subsection{From stripes to islands of shape coexistence} 

The stripes  7-8, 17-20, 34-40, 59-70, 96-112, 146-168 determined in the previous subsection represent a necessary condition for the appearance of SC, but not a sufficient one. Further work is needed in order to narrow down the stripes into islands. A step in this direction has been taken within covariant density functional theory  \cite{Ring1996,Bender2003,Vretenar2005,Meng2006,Niksic2011,Meng2015,Liang2015}, using the DDME2 functional 
\cite{LalazissisPRC71} and the code of Ref. \cite{Niksic}. A systematic search has been made  \cite{CDFTPLB,CDFTPRC} for particle-hole excitations, which are thought for a long time of being the microscopic mechanism behind SC. Indeed specific islands of SC have been located  \cite{CDFTPLB,CDFTPRC} around the proton shell closures Z=82 and Z=50, in which proton particle-hole excitations are caused by the neutrons, therefore characterizing these cases as ones of neutron-induced shape coexistence. In addition, specific islands of SC have been located  \cite{CDFTPLB,CDFTPRC} around the neutron numbers N=90 and N=60, in which neutron particle-hole excitations are caused by the protons, therefore characterizing these cases as ones of proton-induced shape coexistence. Furthermore, an island of SC has been found around $Z=N=40$, in which both the proton-induced and neutron-induced mechanisms are present simultaneously.  All these islands are consistent with the stripes of the dual shell mechanism within the proxy-SU(3) symmetry, as well as with the empirical islands reviewed in Fig. 8 of Ref. \cite{HW}.

Further corroboration of the predictions of the dual shell mechanism has been provided by various relativistic microscopic calculations \cite{Sharma988,Kumar1001,Thakur52,Thakur1014,Yang103,Mennana96}, as well as by calculations using the Bohr Hamiltonian \cite{Mennana96,Mennana105} and the IBM \cite{Hosseinnezhad1028}.  

It should be noticed that the above predictions of specific islands of SC  are based on the particle-hole excitation mechanism. It could be possible that some other microscopic mechanism creates SC in regions of the nuclear chart outside the islands predicted here. This can be the subject of further investigation. 

\subsection{Multiple shape coexistence}

Recent experimental evidence exists that multiple shape coexistence of up to four bands can be seen in certain nuclei \cite{Garrett4,Garrett5}, while it is also predicted theoretically in others \cite{Yang103}.  Multiple shape coexistence can occur within the dual shell mechanism described above, since the protons can follow either the SO or the HO scheme, and so can independently do the neutrons. As a result, four different irreps, based on the proton-neutron combinations SO-SO, SO-HO, HO-SO, HO-HO can occur in general, giving rise to multiple coexistence of four bands, or three bands in the special case in which equal numbers of valence protons and valence neutrons occupy the same shell. This idea, however, still needs to be tested against experimental evidence.  

\section{Conclusion and outlook}

In this article we have discussed the physical ideas which led to the introduction of the proxy-SU(3) symmetry, the calculations proving its validity and its connection to the shell model framework,
as well as its first successful applications in predicting in a parameter-free way the values of the collective variables $\beta$ and $\gamma$ for even-even nuclei, as well as the dominance of prolate over oblate shapes in the ground states of even-even nuclei, a prolate to oblate shape/phase transition, and the existence of islands on the nuclear chart on which shape coexistence can appear.   
Several directions for further investigations are calling attention, as we shall briefly discuss below.

The proxy-SU(3) symmetry offers the possibility of making parameter-free predictions for B(E2) transition rates. The value of $B(E2; 2_1^+\to 0_1^+)$ is known to be connected to the collective variable $\beta$ \cite{Raman,Pritychenko}, thus it can be determined in a parameter-free way from the $\beta$ values calculated as described in Sec. XIII.B. Since ratios of B(E2)s will only depend on angular momentum coupling coefficients of SO(3) \cite{Edmonds,Varshalovich} and SU(3) \cite{Akiyama14,Akiyama5,Millener,Bahri41,Bahri159,SU3LIB}, parameter-free predictions for all B(E2)s could be obtained in principle. Some first steps in this direction have been taken in \cite{36J,41GC}. 

Based on the experience acquired within the pseudo-SU(3) model \cite{DW1,DW2}, it is expected that the description of nuclear spectra within the proxy-SU(3) symmetry will require the use of third-order and fourth-order operators. In particular, the O(3) symmetry-preserving three-body operator $\Omega$ and four-body operator $\Lambda$ (their mathematical names being the $O_l^0$ and $Q_l^0$ shift operator respectively)\cite{Hughes1,Hughes2,Judd,DeMeyer1} will be needed in order to break the degeneracy between the ground state band (gsb) and the $\gamma_1$ band, which in the proxy-SU(3) approach lie in general within the same SU(3) irrep. This can be seen, for example, in Table VII, where almost all nuclei are characterized by SU(3) irreps with $\mu \geq 2$. Since $K$ takes the values $K=0,2,\dots,\mu$ \cite{IacLie,Elliott1,IA}, the hw irrep will contain both the $K=0$ (gsb) and $K=2$ ($\gamma_1$) bands \cite{Hosseinnezhad1022}. The parameter-free reproduction of the empirical observation that the energy differences between the $\gamma_1$ band and the ground state band decrease as a function of the angular momentum $L$ in deformed nuclei \cite{NPA1009}, with the opposite trend seen in vibrational and $\gamma$-unstable nuclei, should be tested. Some first steps in this direction have been taken in Refs. \cite{38J,EC15}. It should be noticed that the energy scale can be fixed in an even-even nucleus by determining the energy of the first excited state $2_1^+$ from the value of the  $B(E2; 2_1^+\to 0_1^+)$, determined in a parameter-free way, as described in section VIII.B. This can be achieved through the microscopically derived \cite{Jolos820,Shirokova105} Grodzins relation \cite{Grodzins2}, connecting the energy of the $2_1^+$ state and $B(E2; 2_1^+\to 0_1^+)$.  

Nuclear binding energies and nucleon separation energies are basic nuclear structure quantities, for which extended experimental data \cite{AME2016} and theoretical predictions \cite{LalazissisADNDT71,Fossion2002,FRDM2012} exist. It is an interesting project to examine the degree at which proxy-SU(3) is able to predict these quantities, preferably in a parameter-free way. Some first steps in this direction have been taken in Refs. \cite{40J,42J}. The calculation of two-neutron separation energies is of particular interest, since the recently discovered \cite{Couture96,Couture104} connection between them and the neutron capture cross sections, which are essential for understanding the astrophysical $s$ and $r$ processes.  

\bigskip 
\section*{Acknowledgements} 

Support by the Bulgarian National Science Fund (BNSF) under Contract No. KP-06-N48/1  is gratefully acknowledged.


\begin{table}[htb] \label{Tab1}
\centering
\caption{Federman--Pittel \cite{FP1,FP2,FP3} pairs of orbitals playing a leading role in the development of nuclear deformation in different mass regions. On the left part of the table are shown the pairs contributing in the beginning of the relevant shell, while the pairs on the right become important further within the shell. Adapted from Ref. \cite{41J}. See section III for further discussion.} 
\begin{tabular}{ r r r r r   }
\hline\noalign{\smallskip}
 &  protons  & neutrons &      protons  &     neutrons  \\ 
\noalign{\smallskip}\hline\noalign{\smallskip}
light       & 1d$_{5/2}$     & 1d$_{3/2}$    &    1d$_{5/2}$      &  1f$_{7/2}$        \\
intermediate& 1g$_{9/2}$     & 1g$_{7/2}$    &    1g$_{9/2}$      &  1h$_{11/2}$       \\
rare earths & 1h$_{11/2}$    & 1h$_{9/2}$    &    1h$_{11/2}$     &  1i$_{13/2}$       \\
actinides   & 1i$_{13/2}$    & 1i$_{11/2}$   &    1i$_{13/2}$     &  1j$_{15/2}$       \\

\noalign{\smallskip}\hline
\end{tabular}
\end{table}

\newpage 

\begin{turnpage}

\begingroup

\squeezetable 

\begin{table}[htb] \label{Tab2}

\caption{Matrix elements of the Nilsson Hamiltonian in the 82--126 neutron shell (upper part)
and in the full pfh neutron shell (lower part). In order to demonstrate the quality of the proxy-SU(3) approximation through comparison of the two sets, matrix elements in the lower part of the table, which differ from their counterparts in the upper part, are shown in boldface. The matrix elements have been calculated for deformation $\epsilon=0.3$ and are shown in units of $\hbar \omega_0$. Adapted from Ref. \cite{proxy1}. See section V for further discussion.}

\bigskip

\begin{tabular}{ r r r r r r r r r r r r r r r r | r r r r r r r }
                  &
${1\over 2}$[501] & ${1\over 2}$[521] & ${3\over 2}$[512] & ${1\over 2}$[510] & ${3\over 2}$[501] & ${5\over 2}$[503] & ${1\over 2}$[541] & 
${3\over 2}$[532] & ${5\over 2}$[523] & ${7\over 2}$[514] & ${1\over 2}$[530] & ${3\over 2}$[521] & ${5\over 2}$[512] & ${7\over 2}$[503] & 
${9\over 2}$[505] & ${1\over 2}$[660] & ${3\over 2}$[651] & ${5\over 2}$[642] & ${7\over 2}$[633] & ${9\over 2}$[624] & 
${11\over 2}$[615] & ${13\over 2}$[606] 
\\

\hline

1/2[501] & 7.44 & 0.19 & 0 & 0.16 & 0  & 0  & 0  & 0  & 0  & 0  & 0  & 0  & 0  & 0 & 0 & 0  & 0 & 0 & 0 & 0 &  0 & 0 \\
1/2[521] &  & 6.46 & 0 & $-0.18$ & 0  & 0  & 0.26  & 0  & 0  & 0  & 0.22  & 0  & 0  & 0 & 0 & 0  & 0 & 0 & 0 & 0 & 0 & 0 \\
3/2[512] &  &  & 6.88 & 0 & $-0.13$  & 0  & 0  & 0.23  & 0  & 0  & 0  & 0.22 & 0  & 0 & 0 & 0  & 0 & 0 & 0 & 0 & 0 & 0 \\
1/2[510] & &  &  & 6.86 & 0  & 0  & 0  & 0  & 0  & 0  & 0.26  & 0  & 0  & 0 & 0 & 0  & 0 &  0 & 0 & 0 & 0 & 0 \\
3/2[501] &    &    &    &    & 7.31 & 0 & 0 & 0 & 0 & 0 & 0 & 0.19 & 0 & 0    & 0    & 0 & 0 & 0   & 0 & 0 & 0   & 0 \\
5/2[503] &     &     &     &     &  & 7.35 & 0 & 0 & 0.15 & 0 & 0 & 0 & 0.18 & 0    & 0    & 0 & 0 & 0   & 0 & 0 & 0   & 0 \\
1/2[541] &   &  &   &   &  & & 5.92 & 0 & 0 & 0  & $-0.18$ & 0 & 0 & 0    & 0    & 0 & 0 & 0   & 0 & 0 & 0   & 0 \\
3/2[532] &   &   &    &    &  &  &  & 6.12 & 0 & 0 & 0  & $-0.16$ & 0 & 0    & 0    & 0 &  0 & 0   & 0 & 0 & 0   & 0 \\
5/2[523] &  &  &  &  &  &  & &  & 6.38 & 0 & 0 & 0  & $-0.13$ & 0    & 0    & 0 & 0 & 0   & 0 & 0 & 0   & 0 \\
7/2[514] &  &  &  &   &  &  &  &  &  & 6.69 & 0 & 0 & 0 & $-0.09$ & 0 & 0 & 0 & 0   & 0 & 0 & 0   & 0 \\
1/2[530] &  & &    &  &  &  &  &  &  &  & 6.10 & 0 & 0 & 0    & 0    & 0 & 0 & 0   & 0 & 0 & 0   & 0 \\
3/2[521] & & &  &   &  & &  & &  &  &  & 6.34 & 0 & 0    & 0    & 0 & 0 & 0   & 0 & 0 & 0   & 0 \\
5/2[512] & &   &  &  &  &  &  &  &  &  &  &  & 6.63 & 0    & 0    & 0 & 0 & 0   & 0 & 0 & 0   & 0 \\
7/2[503] &  &  &  &  &   &   &   &   &   &   &   &   &   & 6.97 & 0 & 0  & 0 & 0 & 0 & 0 & 0 & 0 \\
9/2[505] &  &  &  &  &   &  &  &  &  &  &  &  & &  & 7.05 & 0  & 0 & 0 & 0 & 0 & 0 & 0 \\
 
 \hline

1/2[660] &   &   &  & &  &  &  &  &  &  &  &  &  &    &     & 6.70 & 0 & 0   & 0 & 0 & 0   & 0 \\
3/2[651] &  &   &  &  &  &  &  &  &  &  &  &  &  &     &     & & 6.67 & 0   & 0 & 0 & 0   & 0 \\
5/2[642] & &  &  &  &   &   &   &   &   &   &   &   &   &  &  &   &  & 6.69 & 0 & 0 & 0 & 0 \\
7/2[633] &  &  &    &  &  &  &  &  &  &  &  &  &  &  & & &  &    & 6.77 & 0 & 0   & 0 \\
9/2[624] &  &  & &  &  &  &  &  &  &  &  &  &  &     &  &  &  &  &  & 6.90 & 0   & 0 \\
11/2[615] &  &  &  &  &   &   &   &   &   &   &   &   &   &  &  &   &  &  &  &  & 7.08 & 0 \\
13/2[606] &  &  &  &  &   &   &   &   &   &   &   &   &   &  &  &  &  &    &  &  &    & 7.32 \\
 
 \hline
\end{tabular}

\vskip 0.3cm 

\begin{tabular}{ r r r r r r r r r r r r r r r r | r r r r r r }
                  &
${1\over 2}$[501] & ${1\over 2}$[521] & ${3\over 2}$[512] & ${1\over 2}$[510] & ${3\over 2}$[501] & ${5\over 2}$[503] & ${1\over 2}$[541] & 
${3\over 2}$[532] & ${5\over 2}$[523] & ${7\over 2}$[514] & ${1\over 2}$[530] & ${3\over 2}$[521] & ${5\over 2}$[512] & ${7\over 2}$[503] & 
${9\over 2}$[505] & ${1\over 2}$[550] & ${3\over 2}$[541] & ${5\over 2}$[532] & ${7\over 2}$[523] & ${9\over 2}$[514] & 
${11\over 2}$[505] 
\\

\hline

1/2[501] & 7.44 & 0.19 & 0 & 0.16 & 0  & 0  & 0  & 0  & 0  & 0  & 0  & 0  & 0  & 0 & 0 & 0  & 0 & 0 & 0 & 0 &  0  \\
1/2[521] &  & 6.46 & 0 & $-0.18$ & 0  & 0  & 0.26  & 0  & 0  & 0  & 0.22  & 0  & 0  & 0 & 0 & 0  & 0 & 0 & 0 & 0 & 0  \\
3/2[512] &  &  & 6.88 & 0 & $-0.13$  & 0  & 0  & 0.23  & 0  & 0  & 0  & 0.22 & 0  & 0 & 0 & 0  & 0 & 0 & 0 & 0 & 0  \\
1/2[510] & & &  & 6.86 & 0  & 0  & 0  & 0  & 0  & 0  & 0.26  & 0  & 0  & 0 & 0 & 0  & 0 &  0 & 0 & 0 & 0 \\
3/2[501] &  & &  &  & 7.31 & 0 & 0 & 0 & 0 & 0 & 0 & 0.19 & 0 & 0    & 0    & 0 & 0 & 0   & 0 & 0 & 0  \\
5/2[503] &  &  &  &  &  & 7.35 & 0 & 0 & 0.15 & 0 & 0 & 0 & 0.18 & 0    & 0    & 0 & 0 & 0   & 0 & 0 & 0  \\
1/2[541] &  &   &   &   &  &  & 5.92 & 0 & 0 & 0  & $-0.18$ & 0 & 0 & 0    & 0    & {\bf 0.20} & 0 & 0   & 0 & 0 & 0  \\
3/2[532] &  &  &   &     &  &  & & 6.12 & 0 & 0 & 0  & $-0.16$ & 0 & 0    & 0    & 0 &  {\bf 0.25} & 0   & 0 & 0 & 0  \\
5/2[523] & & &  &  & &  &  &  & 6.38 & 0 & 0 & 0  & $-0.13$ & 0    & 0    & 0 & 0 & {\bf 0.27}   & 0 & 0 & 0 \\
7/2[514] &   &  &  &  & &  &  &  & & 6.69 & 0 & 0 & 0 & $-0.09$ & 0 & 0 & 0 & 0   & {\bf 0.25} & 0 & 0 \\
1/2[530] &  & &   &   &  &  &  & &  &  & 6.10 & 0 & 0 & 0    & 0    & {\bf 0.24} & 0 & 0   & 0 & 0 & 0 \\
3/2[521] & &  &  &   &  &  &  &  &  &  &  & 6.34 & 0 & 0    & 0    & 0 & {\bf 0.26} & 0   & 0 & 0 & 0  \\
5/2[512] & &  &  &  &  &  &  &  &  &  &  &  & 6.63 & 0    & 0    & 0 & 0 & {\bf 0.23}   & 0 & 0 & 0  \\
7/2[503] &  &  &  &  &   &   &   &   &   &   &   &   &  & 6.97 & 0 & 0  & 0 & 0 & {\bf 0.15} & 0 & 0 \\
9/2[505] &  &  &  &  &   &   &   &   &   &   &   &   &   &  & 7.05 & 0  & 0 & 0 & 0 & {\bf 0.20} & 0 \\
 
 \hline

1/2[550] & & & & & & &  & & &  &  &  &  &    &    & {\bf 6.57} & 0 & 0   & 0 & 0 & 0 \\
3/2[541] & & &  &  &  &  &  &  &  &  &  &  &  &  &  &  & {\bf 6.59} & 0   & 0 & 0 & 0 \\
5/2[532] & &  &  &  &  &  &  &  &   &   &   &   &   &  &  &   &  & {\bf 6.67} & 0 & 0 & 0 \\
7/2[523] & & &  &  & &  &  &  &  &  &  &  &  &   &  &  &  & & {\bf 6.80} & 0 & 0 \\
9/2[514] & & &  & &  &  &  &  &  &  &  &  &  &     &   &  &  &  &  & {\bf 6.98} & 0\\
11/2[505] &  &  &  &  &  &  &  &  &  & &  &  &  &  &  &   &  &  &  &  & {\bf 7.21}   \\

 \hline
\end{tabular}

\end{table}

\endgroup

\end{turnpage}

\newpage


\begin{table*}[htb] \label{Tab3}

\caption{Shell model orbitals of the original spin-orbit like shells are listed next to their proxy-SU(3) counterparts. Orbitals being replaced are indicated in boldface. The sub-shell closure at 14 nucleons has been proposed  in Ref. \cite{Sorlin}. Each proxy-SU(3) shell possesses the symmetry U($\Omega$) with $\Omega={(\mathcal{N}+1)(\mathcal{N}+2)/ 2}$.  Adapted from Ref. \cite{EPJASM}. See Section VI for further discussion.}

\begin{tabular}{cccccc}
\noalign{\smallskip}\hline\noalign{\smallskip}
spin-orbit    &                    &                &                            & proxy-SU(3)    &  3D-HO        \\          
magic numbers & original orbitals & proxy orbitals &proxy  $U(\Omega)$ symmetry &  magic numbers & magic numbers \\
 \noalign{\smallskip}\hline\noalign{\smallskip}
 
6-14 & $1p^{1/2}_{\pm 1/2}$ & $1p^{1/2}_{\pm 1/2}$ & $U(3)$ & 6-12 & 2-8 \smallskip\\
&  $\bf 1d^{5/2}_{\pm 1/2,\pm 3/2}$&  $\bf 1p^{3/2}_{\pm 1/2,\pm 3/2}$ & & &  \smallskip\\
 & $\bf 1d^{5/2}_{\pm 5/2}$ & - & & \bigskip\\
 
 14-28 & $2s^{1/2}_{\pm 1/2}$ & $2s^{1/2}_{\pm 1/2}$ &  $U(6)$ & 14-26 & 8-20 \smallskip\\
 & $1d^{3/2}_{\pm 1/2,\pm 3/2}$ & $1d^{3/2}_{\pm 1/2,\pm 3/2}$ & & & \smallskip\\
 & $\bf 1f^{7/2}_{\pm 1/2,\pm 3/2,\pm 5/2}$ & $\bf 1d^{5/2}_{\pm 1/2,\pm 3/2,\pm 5/2}$ & & & \smallskip\\
& $\bf 1f^{7/2}_{\pm 7/2}$ & - & & & \bigskip\\

28-50 & $2p^{1/2}_{\pm 1/2}$ & $2p^{1/2}_{\pm 1/2}$ &  $U(10)$ & 28-48 & 20-40 \smallskip\\
& $2p^{3/2}_{\pm 1/2,\pm 3/2}$ &  $2p^{3/2}_{\pm 1/2,\pm 3/2}$ & & & \smallskip\\
& $1f^{5/2}_{\pm 5/2, \pm3/2,\pm 1/2}$ & $1f^{5/2}_{\pm 5/2, \pm3/2,\pm 1/2}$ & & & \smallskip\\
& $\bf 1g^{9/2}_{\pm 1/2,..., \pm 7/2}$ & $\bf 1f^{7/2}_{\pm 1/2,..., \pm 7/2}$ & & & \smallskip\\
& $\bf 1g^{9/2}_{\pm 9/2}$ & - & & & \bigskip\\

50-82  & $3s^{1/2}_{\pm 1/2}$ &  $3s^{1/2}_{\pm 1/2}$  & $U(15)$ & 50-80 & 40-70 \smallskip\\
 & $2d^{3/2}_{\pm 1/2,\pm 3/2}$ & $2d^{3/2}_{\pm 1/2,\pm 3/2}$ & & & \smallskip\\
& $2d^{5/2}_{\pm 1/2,...,\pm 5/2}$ &  $2d^{5/2}_{\pm 1/2,...,\pm 5/2}$ & & & \smallskip\\
 & $1g^{7/2}_{\pm 1/2,...,\pm 7/2}$ & $1g^{7/2}_{\pm 1/2,...,\pm 7/2}$ & & & \smallskip\\
 & $\bf 1h^{11/2}_{\pm 1/2,...,\pm 9/2}$ & $\bf 1g^{9/2}_{\pm 1/2,...,\pm 9/2}$ & & & \smallskip\\
 & $\bf 1h^{11/2}_{\pm 11/2}$ & - & & & \bigskip\\
 
82-126 & $3p^{1/2}_{\pm 1/2}$ & $3p^{1/2}_{\pm 1/2}$ & $U(21)$ & 82-124 & 70-112 \smallskip\\
 & $3p^{3/2}_{\pm 1/2,\pm 3/2}$ & $3p^{3/2}_{\pm 1/2,\pm 3/2}$ & & & \smallskip\\
& $2f^{5/2}_{\pm 1/2,...,\pm 5/2}$ & $2f^{5/2}_{\pm 1/2,...,\pm 5/2}$ & & & \smallskip\\
& $2f^{7/2}_{\pm 1/2,...,\pm 7/2}$ & $2f^{7/2}_{\pm 1/2,...,\pm 7/2}$ & & & \smallskip\\
 & $1h^{9/2}_{\pm 1/2,...,\pm 9/2}$ & $1h^{9/2}_{\pm 1/2,...,\pm 9/2}$ & & & \smallskip\\
 & $\bf 1i^{13/2}_{\pm 1/2,...,\pm 11/2}$ & $\bf 1h^{11/2}_{\pm 1/2,...,\pm 11/2}$ & & & \smallskip\\
 & $\bf 1i^{13/2}_{\pm 13/2}$ & - & & & \bigskip \\
 
126-184 & $4s^{1/2}_{\pm 1/2}$ & $4s^{1/2}_{\pm 1/2}$  & $U(28)$ & 126-182 & 112-168 \smallskip\\
& $3d^{3/2}_{\pm 1/2,\pm 3/2}$ &  $3d^{3/2}_{\pm 1/2,\pm 3/2}$ & & & \smallskip\\
& $3d^{5/2}_{\pm 1/2,...,\pm 5/2}$ & $3d^{5/2}_{\pm 1/2,...,\pm 5/2}$ & & \smallskip\\
& $2g^{7/2}_{\pm 1/2,...,\pm 7/2}$ &  $2g^{7/2}_{\pm 1/2,...,\pm 7/2}$& & & \smallskip\\
& $2g^{9/2}_{\pm 1/2,...,\pm 9/2}$ &  $2g^{9/2}_{\pm 1/2,...,\pm 9/2}$ & & & \smallskip\\
& $1i^{11/2}_{\pm 1/2,...,\pm 11/2}$ & $1i^{11/2}_{\pm 1/2,...,\pm 11/2}$ & & & \smallskip\\
& $\bf 1j^{15/2}_{\pm 1/2,...,\pm 13/2}$ & $\bf 1i^{13/2}_{\pm 1/2,...,\pm 13/2}$ & & & \smallskip\\
& $\bf 1j^{15/2}_{\pm 15/2}$ & - & & & \bigskip\\

\noalign{\smallskip}\hline\noalign{\smallskip}
\end{tabular}
\end{table*}

\newpage

\begin{table*}[htb] \label{Tab4}
\centering
\caption{The table proves that Nilsson orbitals possessing the highest total angular momentum $j$ in their shell exhibit a leading shell model eigenvector at three different deformations $\epsilon$.
 The Nilsson orbitals $K[N n_z \Lambda]$ are expanded in the shell model basis $|N l j m_j \rangle$ .  Adapted from Ref. \cite{EPJPSM}. See section VI for further discussion. 
}
\begin{tabular}{ r c r r r r r r  }
\hline\noalign{\smallskip}
${3\over 2}[541]$ & $\epsilon$ & $|N l j m_j \rangle$ & 
$\left| 5 1 {3\over 2} {3\over 2} \right\rangle$ & 
$\left| 5 3 {5\over 2} {3\over 2} \right\rangle$ & 
$\left| 5 3 {7\over 2} {3\over 2} \right\rangle$ & 
$\left| 5 5 {9\over 2} {3\over 2} \right\rangle$ & 
$\left| 5 5 {11\over 2} {3\over 2} \right\rangle$ \\

\noalign{\smallskip}\hline\noalign{\smallskip}
 & 0.05 & & 0.0025 & $-0.0015$ & 0.0641 & $-0.0122$ & 0.9979 \\
 & 0.22 & & 0.0371 & $-0.0286$ & 0.2565 & $-0.0640$ & 0.9633 \\
 & 0.30 & & 0.0601 & $-0.0506$ & 0.3287 & $-0.0922$ & 0.9366 \\

\noalign{\smallskip}\hline
\end{tabular}

\begin{tabular}{ r c  r r r r r r r }
\hline\noalign{\smallskip}
${3\over 2}[651]$ & $\epsilon$ & $|N l j m_j \rangle$ & 
$\left| 6 2 {3\over 2} {3\over 2} \right\rangle$ & 
$\left| 6 2 {5\over 2} {3\over 2} \right\rangle$ & 
$\left| 6 4 {7\over 2} {3\over 2} \right\rangle$ & 
$\left| 6 4 {9\over 2} {3\over 2} \right\rangle$ & 
$\left| 6 6 {11\over 2} {3\over 2} \right\rangle$ &
$\left| 6 6 {13\over 2} {3\over 2} \right\rangle$ \\

\noalign{\smallskip}\hline\noalign{\smallskip}

 & 0.05 & & $-0.0002$ & 0.0046 & $-0.0013$ & 0.0821 & $-0.0086$ & 0.9966  \\
 & 0.22 & & $-0.0100$ & 0.0711 & $-0.0278$ & 0.3240 & $-0.0469$ & 0.9418  \\
 & 0.30 & & $-0.0207$ & 0.1149 & $-0.0509$ & 0.4091 & $-0.0687$ & 0.9010  \\

\noalign{\smallskip}\hline
\end{tabular}

\end{table*}

\begin{table*}[htb] \label{Tab5}
\centering
\caption{The table proves that Nilsson orbitals not possessing the highest total angular momentum $j$ in their shell exhibit a leading shell model eigenvector only at small deformation $\epsilon$, while at higher deformations several shell model eigenvectors make sizable contributions. The Nilsson orbitals $K[N n_z \Lambda]$ are expanded in the shell model basis $|N l j m_j \rangle$.  Adapted from Ref. \cite{EPJPSM}.  See section VI for further discussion. 
} 
\begin{tabular}{r c  r r r r r r  }
\hline\noalign{\smallskip}
${1\over 2}[431]$ & $\epsilon$ & $|N l j m_j \rangle$ & 
$\left| 4 0 {1\over 2} {1\over 2} \right\rangle$ & 
$\left| 4 2 {3\over 2} {1\over 2} \right\rangle$ & 
$\left| 4 2 {5\over 2} {1\over 2} \right\rangle$ & 
$\left| 4 4 {7\over 2} {1\over 2} \right\rangle$ & 
$\left| 4 4 {9\over 2} {1\over 2} \right\rangle$ \\

\noalign{\smallskip}\hline\noalign{\smallskip}
 & 0.05 & & $-0.0213$ & 0.1254 & $-0.0702$ & 0.9893 & 0.0127 \\
 & 0.22 & & $-0.2248$ & 0.4393 & $-0.2791$ & 0.8057 & 0.1717 \\
 & 0.30 & & $-0.2630$ & 0.5003 & $-0.2458$ & 0.7447 & 0.2559 \\

\noalign{\smallskip}\hline
\end{tabular}

\begin{tabular}{ r c r r r r r r r }
\hline\noalign{\smallskip}
${1\over 2}[541]$ & $\epsilon$ & $|N l j m_j \rangle$ & 
$\left| 5 1 {1\over 2} {1\over 2} \right\rangle$ & 
$\left| 5 1 {3\over 2} {1\over 2} \right\rangle$ & 
$\left| 5 3 {5\over 2} {1\over 2} \right\rangle$ & 
$\left| 5 3 {7\over 2} {1\over 2} \right\rangle$ & 
$\left| 5 5 {9\over 2} {1\over 2} \right\rangle$ &
$\left| 5 5 {11\over 2} {1\over 2} \right\rangle$ \\

\noalign{\smallskip}\hline\noalign{\smallskip}

 & 0.05 & & $-0.0200$ & 0.1770 & $-0.0295$ & 0.9780 & $-0.0446$ & $-0.0944$  \\
 & 0.22 & & $-0.2492$ & 0.4619 & $-0.3768$ & 0.5550 & $-0.4161$ & $-0.3185$  \\
 & 0.30 & & $-0.3121$ & 0.4331 & $-0.4829$ & 0.3430 & $-0.4789$ & $-0.3671$  \\

\noalign{\smallskip}\hline
\end{tabular}

\end{table*}


\begin{table*}[htb]\label{Tab6}

\caption{Highest weight SU(3) irreps (labeled by hw) for U(n), n=6, 10, 15, 21, 28, 36, and highest $C_2$ irreps (labeled by C) for n=6, 10, 15, 21. Highest weight (hw) irreps differing from their highest $C_2$ counterparts are shown in boldface. The hw irreps alone can be obtained from an analytic formula \cite{Kotaformula}. More irreps can be obtained through the codes of Refs. \cite{Lopezcode,Langr,Alex} and have been  presented in Ref.  \cite{AssimakisMSc}. Adapted from Ref. \cite{37J}. See section VII for further discussion.}

\bigskip
\begin{tabular}{ r l r r r r r r r r r r } 

\hline

\hline
   &             & 8-20 & 8-20 & 28-50 & 28-50& 50-82 & 50-82 & 82-126&82-126 &126-184&184-258\\
   &             & sd   &  sd  & pf    &  pf  & sdg   &  sdg  &  pfh  & pfh   & sdgi  & pfhj  \\
M  & irrep       & U(6) & U(6) & U(10) & U(10)& U(15) & U(15) & U(21) & U(21) & U(28) & U(36) \\
   &             & hw   & C    & hw    &  C   & hw    &  C    & hw    &   C   & hw    & hw    \\
 0 &             &(0,0) &(0,0) &(0,0)  &(0,0) &(0,0)  &(0,0)  &(0,0)  &(0,0)  & (0,0) & (0,0) \\  
 1 & [1]         &(2,0) &(2,0) & (3,0) &(3,0) & (4,0) &(4,0)  & (5,0) &(5,0)  & (6,0) & (7,0) \\
 2 & [2]         &(4,0) &(4,0) & (6,0) &(6,0) & (8,0) &(8,0)  &(10,0) &(10,0) &(12,0) & (14,0)\\
 3 & [21]        &(4,1) &(4,1) & (7,1) &(7,1) &(10,1) &(10,1) &(13,1) &(13,1) &(16,1) & (19,1) \\
 4 & [$2^2$]     &(4,2) &(4,2) & (8,2) &(8,2) &(12,2) &(12,2) &(16,2) &(16,2) &(20,2) & (24,2)\\
 5 & [$2^2$1]    &(5,1) &(5,1) &(10,1) &(10,1)&(15,1) &(15,1) &(20,1) &(20,1) &(25,1) &(30,1) \\
 6 & [$2^3$]     &(6,0) & (0,6)&(12,0) &(12,0)&(18,0) &(18,0) &(24,0) &(24,0) &(30,0) & (36,0)\\
 7 & [$2^3$1]  &{\bf(4,2)}&(1,5)&(11,2)&(11,2)&(18,2) &(18,2) &(25,2) &(25,2) &(32,2) &(39,2)\\
 8 & [$2^4$]     &(2,4) & (2,4)&(10,4) &(10,4)&(18,4) &(18,4) &(26,4) &(26,4) &(34,4) & (42,4)\\
 9 & [$2^4$1]    &(1,4) & (1,4)&(10,4) &(10,4)&(19,4) &(19,4) &(28,4) &(28,4) &(37,4) &(46,4)\\
10 & [$2^5$]     &(0,4) & (0,4)&(10,4) &(4,10)&(20,4) &(20,4) &(30,4) &(30,4) &(40,4) & (50,4)\\
11 & [$2^5$1]   &(0,2)&(0,2)&{\bf(11,2)}&(4,10)&(22,2)&(22,2) &(33,2) &(33,2) &(44,2) &(55,2) \\
12 & [$2^6$]    &(0,0)&(0,0)&{\bf(12,0)}&(4,10)&(24,0)&(24,0) &(36,0) &(36,0) &(48,0) & (60,0)\\
13 & [$2^6$1]   &     &     &{\bf(9,3)} &(2,11)&(22,3)&(22,3) &(35,3) &(35,3) &(48,3) &(61,3)\\
14 & [$2^7$]    &     &     &{\bf(6,6)} &(0,12)&(20,6)&(20,6) &(34,6) &(34,6) &(48,6) &(62,6) \\
15 & [$2^7$1]   &     &     &{\bf(4,7)} &(1,10)&(19,7)& (7,19)&(34,7) &(34,7) &(49,7) &(64,7) \\
16 & [$2^8$]    &  &      & (2,8) & (2,8)&{\bf(18,8)} & (6,20)&(34,8) &(34,8) &(50,8) &(66,8)  \\
17 & [$2^8$1]   &  &      & (1,7) & (1,7)&{\bf(18,7)} & (3,22)&(35,7) &(35,7) &(52,7) &(69,7)\\
18 & [$2^9$]    &  &      & (0,6) & (0,6)&{\bf(18,6)} & (0,24)&(36,6) &(36,6) &(54,6) &(72,6) \\
19 & [$2^9$1]   &  &      & (0,3) & (0,3)&{\bf(19,3)} & (2,22)&(38,3) &(38,3) &(57,3) &(76,3) \\
20 & [$2^{10}$] &  &      & (0,0) & (0,0)&{\bf(20,0)} & (4,20)&(40,0) &(40,0) &(60,0) &(80,0) \\
21 & [$2^{10}$1]&  &      &       &      &{\bf(16,4)} & (4,19)&(37,4) & (4,37)&(58,4) &(79,4)\\
22 & [$2^{11}$] &  &      &       & &{\bf(12,8)} & (4,18)&{\bf(34,8)} &(0,40) &(56,8) &(78,8) \\
23 & [$2^{11}$1]&  &      &       & &{\bf(9,10)} & (2,18)&{\bf(32,10)}&(3,38) &(55,10)& (78,10) \\
24 & [$2^{12}$] &  &      &       & &{\bf(6,12)} & (0,18)&{\bf(30,12)}&(6,36) &(54,12) &(78,12) \\
25 & [$2^{12}$1]&  &      &       & &{\bf(4,12)} & (1,15)&{\bf(29,12)}&(7,35) &(54,12) &(79,12) \\
26 & [$2^{13}$] &      &      &   &      &(2,12) & (2,12)&{\bf(28,12)}&(8,34) &(54,12) &(80,12) \\
27 & [$2^{13}$1]&      &      &   &      &(1,10) & (1,10)&{\bf(28,10)}&(7,34) &(55,10) &(82,10)  \\
28 & [$2^{14}$] &      &      &   &      & (0.8) & (0,8) &{\bf(28,8)} &(6,34) &(56,8) &(84,8) \\
29 & [$2^{14}$1]&      &      &   &      & (0,4) & (0,4) &{\bf(29,4)} &(3,35) &{\bf(58,4)} & (87,4)\\
30 & [$2^{15}$] &      &      &   &      & (0,0) & (0,0) &{\bf(30,0)} &(0,36) &{\bf(60,0)}&(90,0) \\
31 & [$2^{15}$1]&      &      &   &      &       &       &{\bf(25,5)} &(2,33) &{\bf(56,5)} &(87,5) \\
32 & [$2^{16}$] &      &      &   &      &       &       &{\bf(20,10)}&(4,30) &{\bf(52,10)} &(84,10)\\
33 & [$2^{16}$1]&      &      &   &      &       &       &{\bf(16,13)}&(4,28) &{\bf(49,13)} &(82,13) \\
34 & [$2^{17}$] &      &      &   &      &       &       &{\bf(12,16)}&(4,26)&{\bf(46,16)} &(80,16)\\
35 & [$2^{17}$1]&      &      &   &      &       &       &{\bf(9,17)} &(2,25)&\bf{(44,17)} &(79,17)\\ 
36 & [$2^{18}$] &      &      &   &      &       &       &{\bf(6,18)} &(0,24)&{\bf(42,18)} &(78,18) \\
37 & [$2^{18}$1]&      &      &   &      &       &       &{\bf(4,17)} &(1,20)&{\bf(41,17)} &\bf{(78,17)} \\
38 & [$2^{19}$] &      &      &        &      &       &       &(2,16) &(2,16)&\bf{(40,16)} &\bf{(78,16)} \\
39 & [$2^{19}$1]&      &      &        &      &       &       &(1,13) &(1,13)&\bf{(40,13)} & \bf{(79,13)} \\
40 & [$2^{20}$] &      &      &        &      &       &       &(0,10) &(0,10)&\bf{(40,10)} &\bf{(80,10)} \\
41 & [$2^{20}$1]&      &      &        &      &       &       &(0,5)  &(0,5) &\bf{ (41,5)} &\bf{(82,5)} \\
42 & [$2^{21}$] &      &      &        &      &       &       &(0,0)  &(0,0) & \bf{(42,0)} &\bf{(84,0)} \\
43 & [$2^{21}$1]&      &      &   &      &       &       &            &      & \bf{(36,6)} & \bf{(79,6)} \\
44 & [$2^{22}$] &      &      &   &      &       &       &            &      &\bf{(30,12)} &\bf{(74,12)} \\
45 & [$2^{22}$1]&      &      &   &      &       &       &            &      &\bf{(25,16)} &\bf{(70,16)}\\
46 & [$2^{23}$] &      &      &   &      &       &       &            &      &\bf{(20,20)} &\bf{(66,20)}\\
47 & [$2^{23}$1]&      &      &   &      &       &       &            &      &\bf{(16,22)} &\bf{(63,22)}\\
48 & [$2^{24}$] &      &      &   &      &       &       &            &      &\bf{(12,24)} &\bf{(60,24)} \\
49 & [$2^{24}$1]&      &      &   &      &       &       &            &      & \bf{(9,24)} &\bf{(58,24)} \\ 
50 & [$2^{25}$] &      &      &   &      &       &       &            &      & \bf{(6,24)} &\bf{(56,24)}  \\
51 & [$2^{25}$1]&      &      &   &      &       &       &            &      &\bf{ (4,22)} &\bf{(55,22)} \\
52 & [$2^{26}$] &      &      &        &      &       &       &       &      & (2,20) &\bf{(54,20)} \\
53 & [$2^{26}$1]&      &      &        &      &       &       &       &      & (1,16) &\bf{(54,16)}  \\
54 & [$2^{27}$] &      &      &        &      &       &       &       &      & (0,12) &\bf{(54,12)}  \\
55 & [$2^{27}$1]&      &      &        &      &       &       &       &      &  (0,6) &\bf{(55,6)}  \\
56 & [$2^{28}$] &      &      &        &      &       &       &       &      &  (0,0) &\bf{(56,0)}  \\
\hline

\end{tabular}

\end{table*} 


\begin{turnpage}

\begingroup

\begin{table}\label{Tab7}

\caption{Highest weight SU(3) irreps for nuclei with 82-126 protons and 126-184 neutrons. Most irreps are prolate, with oblate irreps being underlined. Adapted from Ref. \cite{37J}. See section VIII.A for further discussion.}

\bigskip

\begin{tabular}{ r r r | r r r r r r r r r r r r r r r r r r r}

& & & Rn & Ra & Th & U & Pu & Cm & Cf & Fm & No & Rf & Sg & Hs & Ds & Cn & Fl & Lv & Og &  &  \\
& & $Z$ &86 & 88 & 90 & 92 & 94 & 96 & 98 & 100 & 102 & 104 & 106 & 108 & 110& 112 & 114 & 116 & 118 & 120 & 122  \\
& &$Z_{val}$ & 4 & 6 & 8 & 10 & 12 & 14 & 16 & 18 & 20 & 22 & 24 & 26 & 28 \\
$N$ &$N_{val}$& irrep &(16,2)&(24,0) & (26,4) & (30,4)& (36,0) & (34,6) & (34,8) & (36,6) & (40,0) & (34,8) & (30,12) & (28,12) & (28,8)&(30,0)&(20,10)&(12,16)&(6,18)&(2,16)&(0,10) \\ 

\hline
130& 4&(20,2)  &(36,4)  &(44,2)  &(46,6) &(50,6)    &(56,2) &(54,8)  &(54,10)&(56,8)  &(60,2) &(54,10)&(50,14)&(48,4)&(48,10)&(50,2) &(40,12)&(32,18)&(26,20)&(22,18)&(20,12)\\
132& 6&(30,0)  &(46,2)  &(54,0)  &(56,4) &(60,4)    &(66,0) &(64,6)  &(64,8)  &(66,6) &(70,0) &(64,8) &(60,12)&(58,12)&(58,8) &(60,0) &(50,10) &(42,16) &(36,18)&(32,16)&(30,10)\\
134& 8&(34,4)  &(50,6)  &(58,4)  &(60,8) &(64,8)    &(70,4) &(68,10)&(68,12)&(70,10)&(74,4) &(68,12)&(64,16)&(62,16)&(62,14)&(64,4)&(54,14)&(46,20)&(40,22)&(36,20)&(34,14)\\
136&10&(40,4) &(56,6)  &(64,4)  &(66,8) &(70,8)    &(76,4) &(74,10)&(74,12)&(76,10)&(80,4) &(74,12)&(70,16)&(68,16)&(68,12)&(70,4)&(60,14)&(52,20)&(46,22)&(42,20)&(40,14)\\ 
138&12&(48,0) &(64,2)  &(72,0)  &(74,4) &(78,4)    &(84,0) &(82,6)  &(82,8)  &(84,6) &(88,0) &(82,8)  &(78,12) &(76,12)&(76,8)  &(78,0)&(68,10)&(60,16)&(54,18)&(50,16)&(48,10)\\
140&14&(48,6) &(64,8)  &(72,6) &(74,10)&(78,10)  &(84,6) &(82,12)&(82,14)&(84,12)&(88,6)&(82,14)&(78,18) &(76,18)&(76,14)&(78,6)&(68,16)&(60,22)&(54,24)&(50,22)&(48,16)\\
142&16&(50,8) &(66,10)&(74,8) &(76,12)&(80,12)  &(86,8) &(84,14)&(84,16)&(86,14)&(90,8)&(84,16)&(80,20)&(78,20)&(78,16)&(80,8) &(70,18)&(62,24)&(56,26)&(52,24)&(50,18)\\
144&18&(54,6) &(70,8)  &(78,6) &(80,10)&(84,10) &(90,6) &(88,12) &(88,14)&(90,12)&(94,6)&(88,14)&(84,18)&(82,18)&(82,14)&(84,6) &(74,16)&(66,22)&(60,24)&(56,22)&(54,16)\\
146&20&(60,0) &(76,2)  &(84,0)  &(86,4) &(90,4)    &(96,0) &(94,6)  &(94,8)  &(96,6)&(100,0)&(94,8) &(90,12) &(88,12)&(88,8) &(90,0) &(80,10) &(72,16)&(66,18)&(62,16)&(60,10)\\
148&22&(56,8) &(72,10)&(80,8)  &(82,12)&(86,12) &(92,8) &(90,14)&(90,16)&(92,14)&(96,8)&(90,16)&(86,20)&(84,20)&(84,16)&(86,8)&(76,18) &(68,24)&(62,26)&(58,24)&(56,18)\\
150&24&(54,12)&(70,14)&(78,12)&(80,16)&(84,16)&(90,12)&(88,18)&(88,20)&(90,18)&(94,12)&(88,20)&(84,24)&(82,24)&(82,20)&(84,12)&(74,22)&(66,28)&(60,30)&(56,28)&(54,22)\\
152&26&(54,12)&(70,14)&(78,12)&(80,16)&(84.16))&(90,12)&(88,18)&(88,20)&(90,18)&(94,12)&(88,20)&(84,24)&(82,24)&(82,20)&(84,12)&(74,22)&(66,28)&(60,30)&(56,28)&(54,22)\\
154&28&(56,8)  &(72,10)&(80,8)  &(82,12)&(86,12) &(92,8) &(90,14)&(90,16)&(92,14) &(96,8)  &(90,16)&(86,20)&(84,20)&(84,16)&(86,8) &(76,18) &(68,24)&(62,26)&(58,24)&(56,18)\\
156&30&(60,0)  &(76,2) &(84,0)  &(86,4) &(90,4)   &(96,0)  &(94,6) &(94,8)   &(96,6)  &(100,0) &(94,8)  &(90,12)&(88,12)&(88,8)  &(90,0) &(80,10) &(72,16)&(66,18)&(62,16)&(60,10)\\
158&32&(52,10)&(68,12)&(76,10)&(78,14)&(82,14)&(88,10)&(86,16)&(86,18)&(88,16)&(92,10)&(86,18) &(82,22)&(80,22)&(80,18)&(82,10)&(72,20)&(64,26)&(58,28)&(54,26)&(52,20)\\
160&34&(46,16)&(62,18)&(70,16)&(72,20)&(76,20)&(82,16)&(80,22)&(80,24)&(82,22)&(86,16)&(80,24)&(76,28)&(74,28)&(74,24) &(76,16)&(66,26)&(58,32)&(52,34)&(48,32)&(46,26)\\
162&36&(42,18)&(58,20)&(66,18)&(68,22)&(72,22)&(78,18)&(76,24)&(76,26)&(78,24)&(82,18)&(76,26)&(72,30)&(70,30)&(70,26)&(72,18)&(62,28)&(54,34)&(48,36)&(44,34)&(42,28)\\
164&38&(40,16)&(56,18) &(64,16)&(66,20)&(70,20)&(76,16)&(74,22)&(74,24)&(76,22)&(80,16)&(74,24)&(70,28)&(68,28)&(68,24)&(70,16)&(60,26)&(52,32)&(46,24)&(42,32)&(40,26)\\
166&40&(40,10)&(56,12) &(64,10)&(66,14)&(70,14)&(76,10)&(74,16)&(74,18)&(76,16)&(80,10)&(74,18)&(70,22)&(68,22)&(68,18)&(70,10)&(60,20)&(52,26)&(46,28)&(42,26)&(40,20)\\
168&42&(42,0)  &(58,2) &(66,0)  &(68,4) &(72,4)   &(78,0)  &(76,6)  &(76,8) &(78,6)  &(82,0)  &(76,8)  &(72,12) &(70,12)&(70,8)  &(72,0)  &(62,10)&(54,16)&(48,18)&(44,16)&(42,10)\\
170&44&(30,12)&(46,14)&(54,12)&(56,16)&(60,16) &(66,12)&(64,18)&(64,20)&(66,18)&(70,12)&(64,20)&(60,24)&(58,24)&(58,20)&(60,12)&(50,22)&(42,28)&(36,30)&(32,28)&(30,22)\\
172&46&(20,20)&(36,22)&(44,20)&(46,24)&(50,24) &(56,20)&(54,26)&(54,28)&(56,26)&(60,20)&(54,28)&(50,32)&(48,32)&(48,28)&(50,20)&(40,30)
&$\underline{(32,36)}$&$\underline{(26,38)}$&$\underline{(22,36}$)&$\underline{(20,30)}$\\
174&48&(12,24)&(28,26)&(36,24)&(38,28)&(42,28) &(48,24)&(46,30)&(46,32)&(48,30)&(52,24)&(46,32)&(42,36)&(40,36)&(40,32)&(42,24)
&$\underline{(32,34}$)&$\underline{(24,40)}$&$\underline{(18,42)}$&
$\underline{(14,40)}$&$\underline{(12,34)}$\\
176&50&(6,24)  &(22,26)&(30,24)&(32,28)&(36,28) &(42,24)&(40,30)&(40,32)&(42,30)&(46,24)&(40,32)&(36,36)
&(34,36)&(34,32)&(36,24)
&$\underline{(26,34)}$&$\underline{(18,40)}$&$\underline{(12,42)}$&$\underline{(8,40)}$  &$\underline{(6,34)}$  \\
178&52&(2,20)  &(18,22)&(26,20)&(28,24)&(32,24) &(38,20)&(36,26)&(36,28)&(38,26)&(42,20)&(36,28)&(32,32)&(30,32)&(30,28)&(32,20)
&$\underline{(22,30)}$
&$\underline{(14,36)}$&$\underline{(8,38)}$  &$\underline{(4,36)}$ 
 &$\underline{(2,30)}$  \\
180&54&(0,12 ) &(16,14)&(24,12)&(26,16)&(30,16)&(36,12)&(34,18&(34,18)&(36,18)&(40,12)&(34,20)&(30,24)&(28,24)&(28,20)&(30,12) &$\underline{(20,22)}$ &$\underline{(12,28)}$ &$\underline{(6,30)}$  &$\underline{(2,28)}$  &$\underline{(0,22)}$\\

\end{tabular}
\end{table}
\endgroup

\end{turnpage}


\begin{table*}
\caption{Using Table VI we report for each nucleon number $M$ the highest weight SU(3) irreps for the spin-orbit (SO) like magic numbers 6, 14, 28, 50, 82, 126 according to the proxy-SU(3) symmetry, as well as the hw SU(3) irreps  for the harmonic oscillator (HO) magic numbers 2, 8, 20, 40, 70, 112, 168 according to the Elliott SU(3) symmetry. Adapted from Ref. \cite{EPJASC}. See section IX for further discussion.}

\begin{center}
\begin{tabular}{c c c c c c | c c c c c c}
$M$ & $(\lambda, \mu)_{SO}$&$(\lambda, \mu)_{HO}$& $M$ & $(\lambda, \mu)_{SO}$&$(\lambda, \mu)_{HO}$ & $M$ & $(\lambda, \mu)_{SO}$&$(\lambda, \mu)_{HO}$& $M$ & $(\lambda, \mu)_{SO}$&$(\lambda, \mu)_{HO}$\\

2 & (0, 0) & (0, 0) & 1 & (0, 0) & (0, 0)   &				       	94& (36, 0) & (30, 12) & 93 & (33, 2)& (32, 10)\\					
4 & (0, 0) & (2, 0) & 3 & (0, 0) & (1, 0)   &				     96& (34, 6) & (28, 12) & 95 & (35, 3) & (29, 12)\\					
6 & (0, 0) & (0, 2) & 5 & (0, 0) & (1, 1)   &				     98& (34, 8) & (28, 8) & 97 & (34, 7) & (28, 10)\\					
8 & (2, 0) & (0, 0) & 7 & (1, 0) & (0, 1)   &				    100& (36, 6) & (30, 0) & 99 & (35, 7) & (29, 4)\\					
10 & (0, 2) & (4, 0) & 9 & (1, 1) & (2, 0)  &					102& (40, 0) & (20, 10) & 101 & (38, 3) & (25, 5)\\					
12 & (0, 0) & (4, 2) & 11 & (0, 1) & (4, 1) &					104& (34, 8) & (12, 16) & 103 & (37, 4) & (16, 13)\\					
14 & (0, 0) & (6, 0) & 13& (0, 0) & (5, 1)  &					106& (30, 12) & (6, 18) & 105 & (32, 10) & (9, 17)\\					
16 & (4, 0) & (2, 4) & 15 & (2, 0) & (4, 2) &					108& (28, 12) & (2, 16) & 107 & (29, 12) & (4, 17)\\					
18 & (4, 2) & (0, 4) & 17 & (4, 1) & (1, 4) &					110& (28, 8) & (0, 10) & 109 & (28, 10) & (1, 13)\\					
20 & (6, 0) & (0, 0) & 19 & (5, 1) & (0, 2) &					112& (30, 0) & (0, 0) & 111 & (29, 4) & (0, 5)\\					
22 & (2, 4) & (6, 0) & 21 & (4, 2) & (3, 0) &					114 & (20, 10) & (12, 0) & 113 & (25, 5) & (6, 0)\\					
24 & (0, 4) & (8, 2) & 23 & (1, 4) & (7, 1) &					116 & (12, 16) & (20, 2) & 115 & (16, 13) & (16, 1)\\					
26 & (0, 0) & (12, 0) & 25 & (0, 2) & (10, 1) &					118 & (6, 18) & (30, 0) & 117 & (9, 17) & (25, 1) \\					
28 & (0, 0) & (10, 4) & 27 & (0, 0) & (11, 2) &					120 & (2, 16) & (34, 4) & 119 & (4, 17) & (32, 2)\\					
30 & (6, 0) & (10, 4) & 29 & (3, 0) & (10, 4) &					122 & (0, 10) & (40, 4) & 121 & (1, 13) & (37, 4)\\					
32 & (8, 2) & (12, 0) & 31 & (7, 1) & (11, 2) &					124 & (0, 0) & (48, 0) & 123 & (0, 5) & (44, 2)\\					
34 & (12, 0) & (6, 6) & 33 & (10, 1) & (9, 3) &					126 & (0, 0) & (48, 6) & 125 & (0, 0) & (48, 3)	\\				
36 & (10, 4) & (2, 8) & 35 & (11, 2) & (4, 7) &					128 & (12, 0) & (50, 8) & 127 & (6, 0) & (49, 7)\\					
38 & (10, 4) & (0, 6) & 37 & (10, 4) & (1, 7) &					130 & (20, 2) & (54, 6) & 129 & (16, 1) & (52, 7)\\					
40 & (12, 0) & (0, 0) & 39 & (11, 2) & (0, 3) &					132 & (30, 0) & (60, 0) & 131 & (25, 1) & (57, 3)\\					
42 & (6, 6) & (8, 0) & 41 & (9, 3) & (4, 0)   &					134 & (34, 4) & (56, 8) & 133 & (32, 2) & (58, 4)\\					
44 & (2, 8) & (12, 2) & 43 & (4, 7) & (10, 1) & 				136 & (40, 4) & (54, 12) & 135 & (37, 4) & (55, 10)\\					
46 & (0, 6) & (18, 0) & 45 & (1, 7) & (15, 1) &					138 & (48, 0) & (54, 12) & 137 & (44, 2) & (54, 12)\\					
48 & (0, 0) & (18, 4) & 47 & (0, 3) & (18, 2) &					140 & (48, 6) & (56, 8) & 139 & (48, 3) & (55, 10)\\					
50 & (0, 0) & (20, 4) & 49 & (0, 0) & (19, 4) &					142 & (50, 8) & (60, 0) & 141 & (49, 7) & (58, 4)\\					
52 & (8, 0) & (24, 0) & 51 & (4, 0) & (22, 2) &					144 & (54, 6) & (52, 10) & 143 & (52, 7) & (56, 5)\\					
54 & (12, 2) & (20, 6) & 53 & (10, 1) & (22, 3) &					146 & (60, 0) & (46, 16) & 145 & (57, 3) & (49, 13)\\					
56 & (18, 0) & (18, 8) & 55 & (15, 1) & (19, 7) &					148 & (56, 8) & (42, 18) & 147 & (58, 4) & (44, 17)\\					
58 & (18, 4) & (18, 6) & 57 & (18, 2) & (18, 7) &					150 & (54, 12) & (40, 16) & 149 & (55, 10) & (41, 17)\\					
60 & (20, 4) & (20, 0) & 59 & (19, 4) & (19, 3) &					152 & (54, 12) & (40, 10) & 151 & (54, 12) & (40, 13)\\					
62 & (24, 0) & (12, 8) & 61 & (22, 2) & (16, 4) &					154 & (56, 8) & (42, 0) & 153 & (55, 10) & (41, 5)\\					
64 & (20, 6) & (6, 12) & 63 & (22, 3) & (9, 10) &					156 & (60, 0) & (30, 12) & 155 & (58, 4) & (36, 6)\\					
66 & (18, 8) & (2, 12) & 65 & (19, 7) & (4, 12) &					158 & (52, 10) & (20, 20) & 157 & (56, 5) & (25, 16)\\					
68 & (18, 6) & (0, 8) & 67 & (18, 7) & (1, 10)  &					160 & (46, 16) & (12, 24) & 159 & (49, 13) & (16, 22)\\					
70 & (20, 0) & (0, 0) & 69 & (19, 3) & (0, 4)   &				162 & (42, 18) & (6, 24) & 161 & (44, 17) & (9, 24)\\					
72&  (12, 8) & (10, 0) & 71 & (16, 4) & (5, 0)  &					164 & (40, 16) & (2, 20) & 163 & (41, 17) & (4, 22)\\					
74& (6, 12) & (16, 2) & 73 & (9, 10) & (13, 1)  &					166 & (40, 10) & (0, 12) & 165 & (40, 13) & (1, 16)\\					
76& (2, 12) & (24, 0) & 75 & (4, 12) & (20, 1)  &					168 & (42, 0) & (0, 0) & 167 & (41, 5) & (0, 6)\\					
78& (0, 8) & (26, 4) & 77 & (1, 10) & (25, 2)   &				170 & (30, 12) & (14, 0) & 169 & (36, 6) & (7, 0)\\					
80& (0, 0) & (30, 4) & 79 & (0, 4) & (28, 4)    &				172 & (20, 20) & (24, 2) & 171 & (25, 16) & (19, 1)\\					
82& (0, 0) & (36, 0) & 81 & (0, 0) & (33, 2)    &				174 & (12, 24) & (36, 0) & 173 & (16, 22) & (30, 1)\\					
84& (10, 0) & (34, 6) & 83 & (5, 0) & (35, 3)   &				176 & (6, 24) & (42, 4) & 175 & (9, 24) & (39, 2)\\					
86& (16, 2) & (34, 8) & 85 & (13, 1) & (34, 7)  &					178 & (2, 20) & (50, 4) & 177 & (4, 22) & (46, 4)\\					
88& (24, 0) & (36, 6) & 87 & (20, 1) & (35, 7)  &					180 & (0, 12) & (60, 0) & 179 & (1, 16) & (55, 2)\\					
90& (26, 4) & (40, 0) & 89 & (25, 2) & (38, 3)  &					182 & (0, 0) & (62, 6) & 181 & (0, 6) & (61, 3)\\					
92& (30, 4) & (34, 8) & 91 & (28, 4) & (37, 4)  &					184 & (0, 0) & (66, 8) & 183 & (0, 0) & (64, 7)					

\end{tabular}
\end{center}
\end{table*}

\newpage


\begin{figure}[htb]\label{Fig1}

\includegraphics[width=75mm]{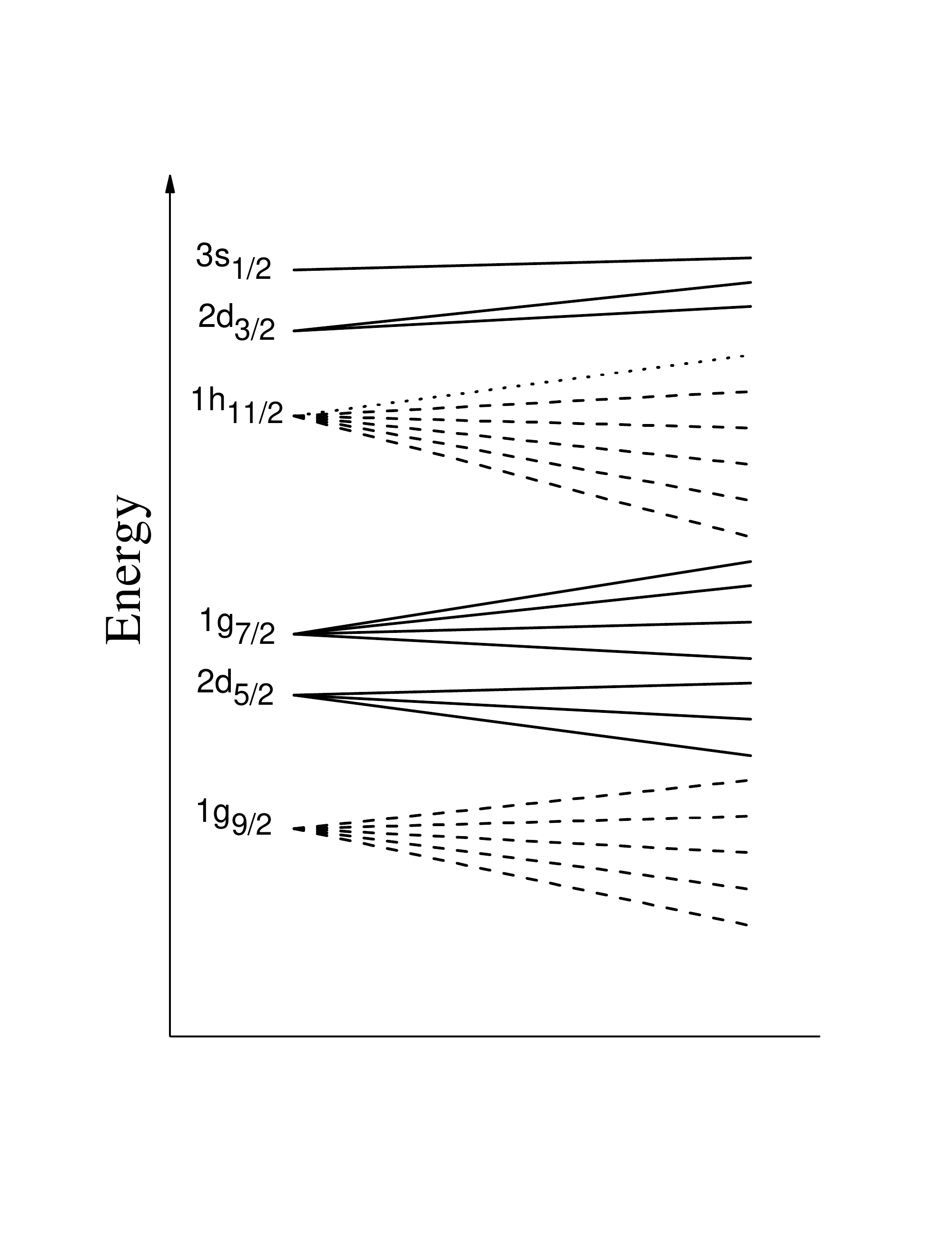}

\caption{Schematic representation of the replacement leading from the 50--82 shell to the proxy sdg shell. Adapted from Ref. \cite{proxy1}. See section IV for further discussion.} 
 
\end{figure}

\begin{figure*}[htb]\label{Fig2}

\includegraphics[width=150mm]{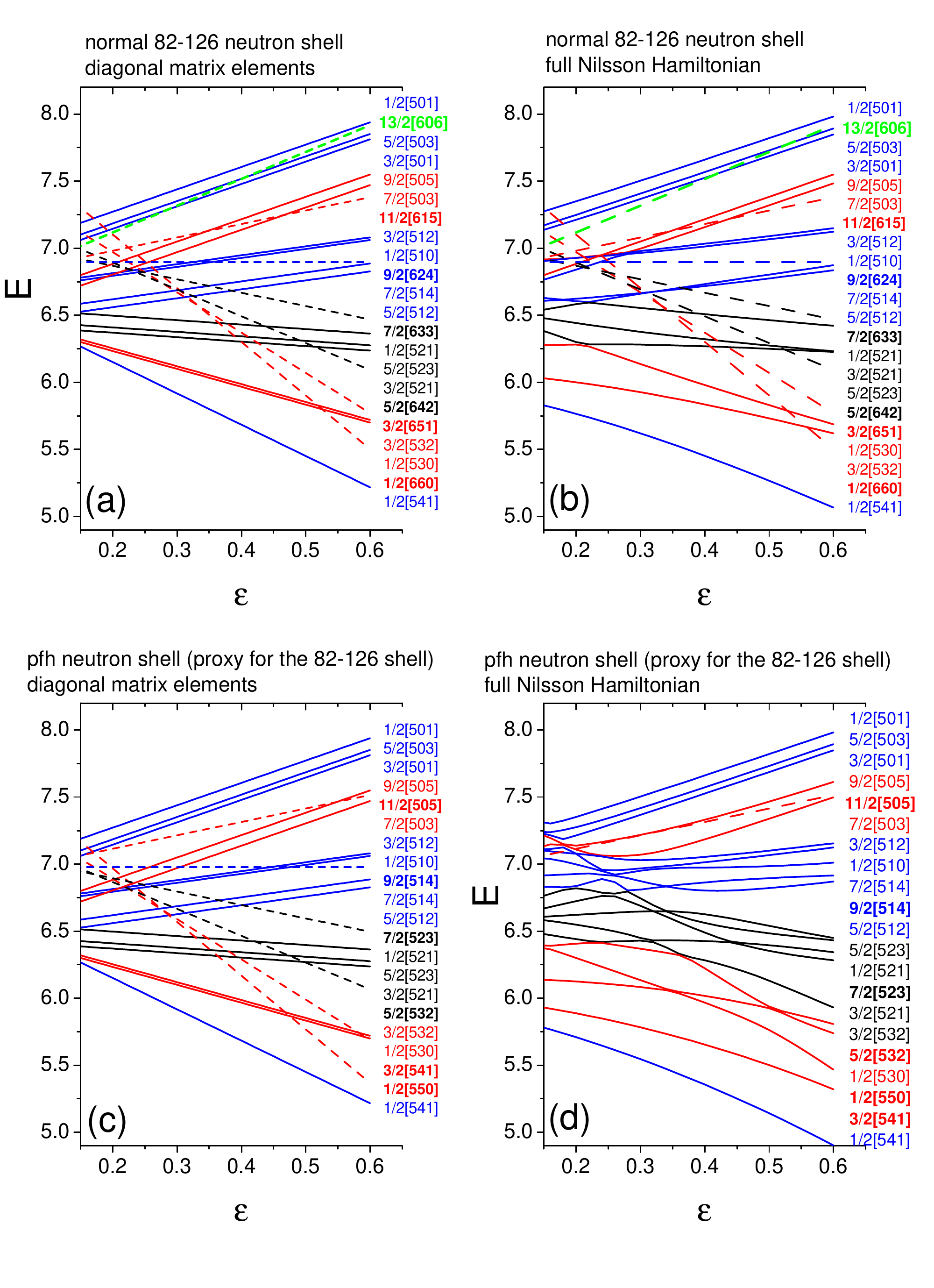}

\caption{The quality of the proxy-SU(3) approximation is demonstrated by showing in the left side of the figure the diagonal matrix elements  of the Nilsson Hamiltonian for the 82-126 (a) and pfh (c) neutron shells and comparing them to the results of the full diagonalization for the same shells, shown in the right side of the figure, in panels (b) and (d) respectively. Orbitals are grouped in color only for guiding the eye.  Intruder orbitals are shown by dashed lines and their labels appear in boldface. All matrix elements are shown as functions of the deformation parameter $\epsilon$, in units of $\hbar \omega_0$. Adapted from Ref. \cite{proxy1}. See section V for further discussion. 
 } 
\end{figure*}


\begin{figure}[htb]\label{Fig3}
\includegraphics[width=75mm]{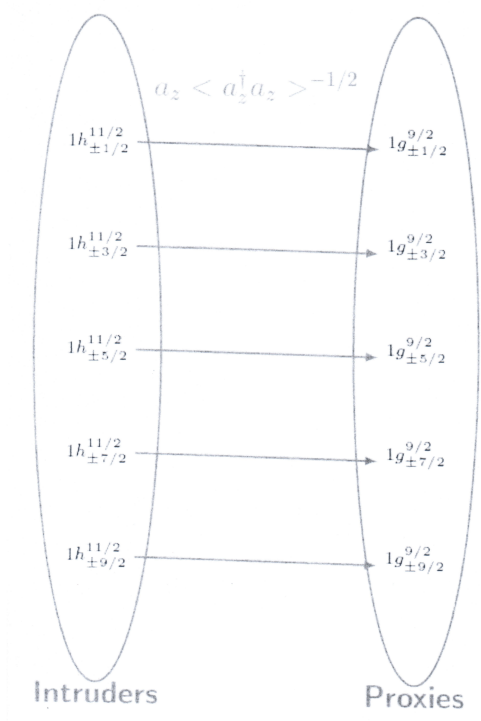}
\caption{Unitary transformation of the intruder orbitals $1h^{11/2}_{m_j}$ (except for the $1h^{11/2}_{\pm 11/2}$) in the 50-82 shell onto the orbitals $1g^{9/2}_{m_j}$.
 Adapted from Ref. \cite{EPJASM}. See section VI for further discussion.}
\end{figure}


\begin{figure*}[htb]\label{Fig5}

{\includegraphics[width=75mm]{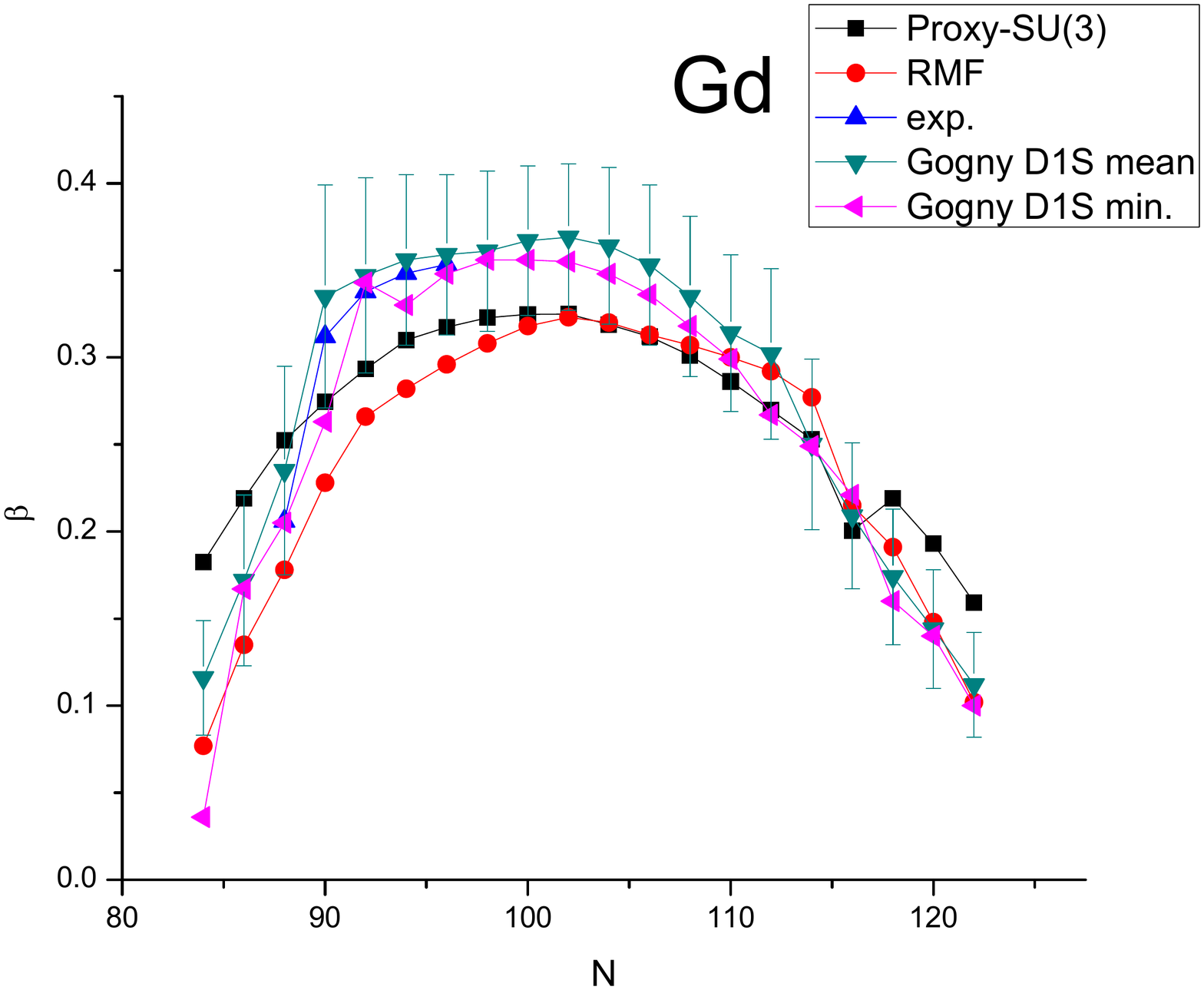}\hspace{5mm}
\includegraphics[width=75mm]{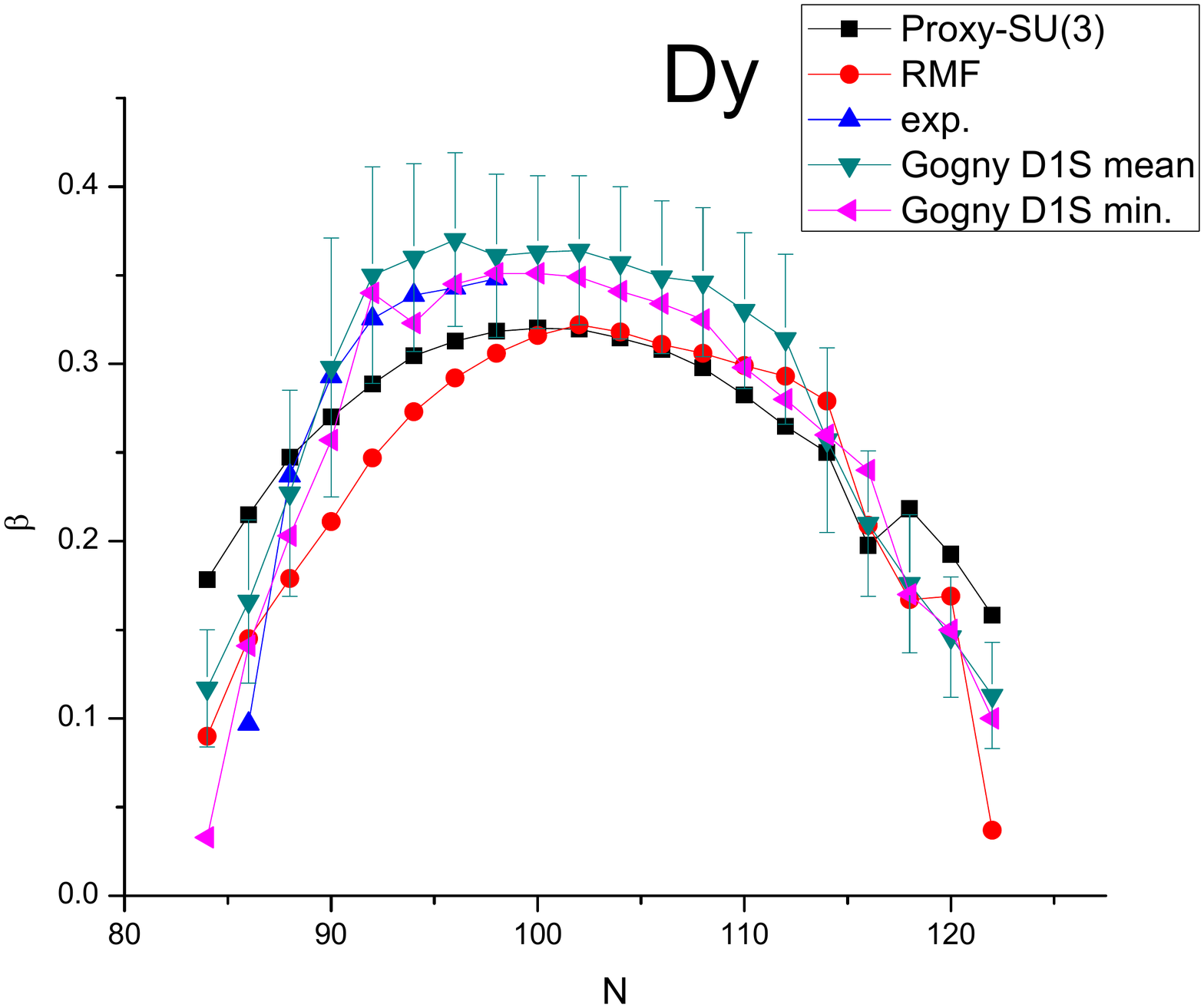}\hspace{5mm}}
{\includegraphics[width=75mm]{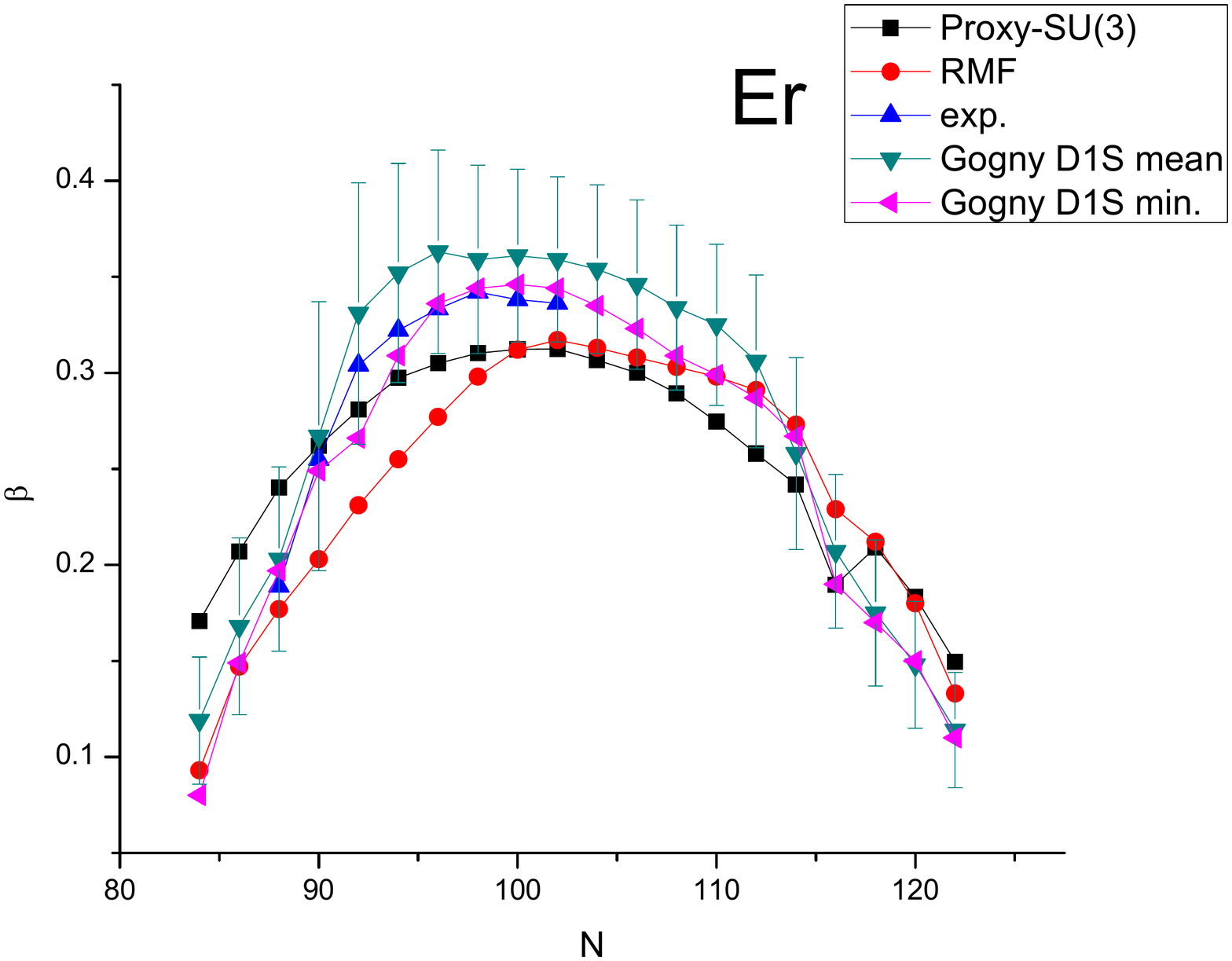}\hspace{5mm}
 \includegraphics[width=75mm]{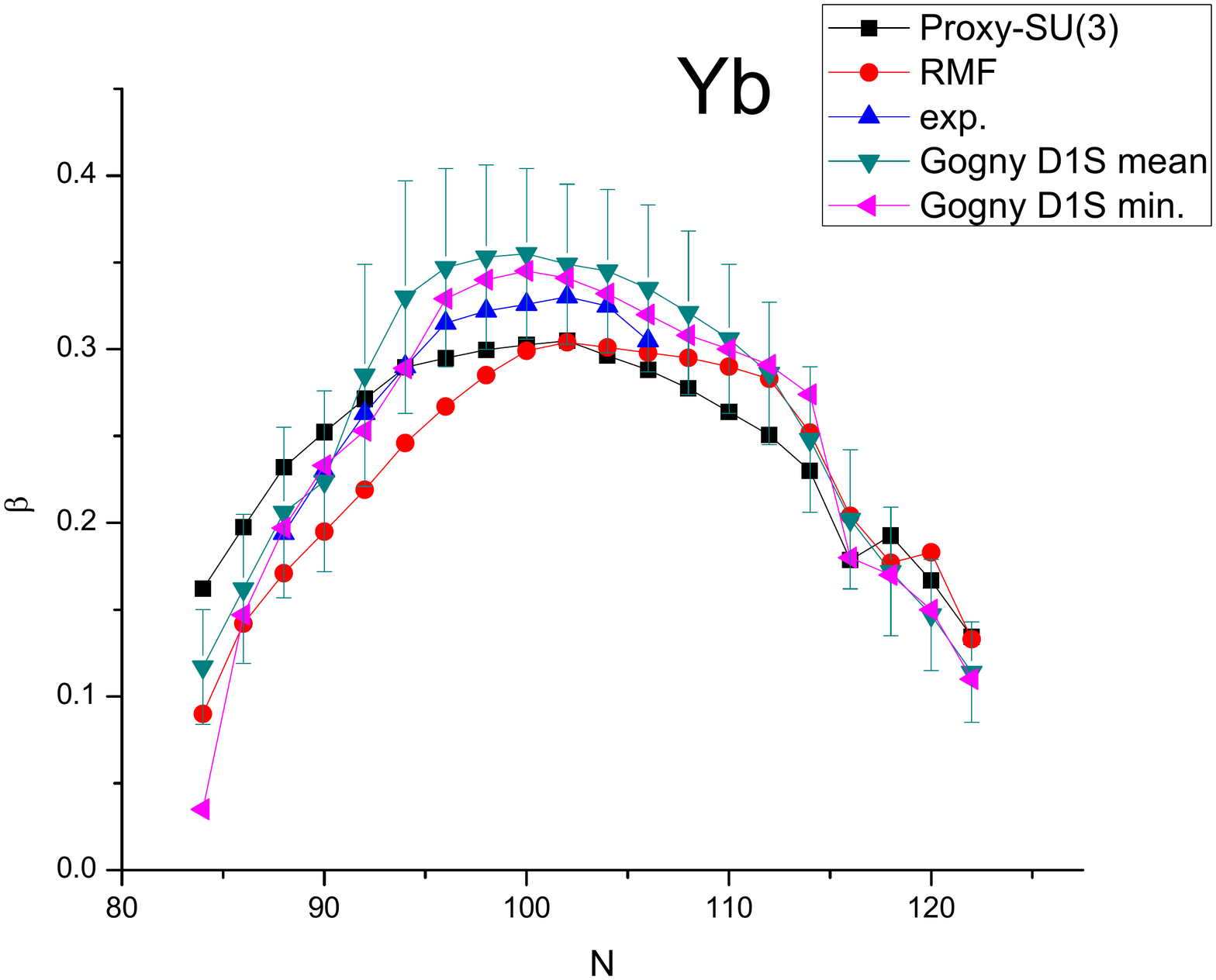}\hspace{5mm}}
{\includegraphics[width=75mm]{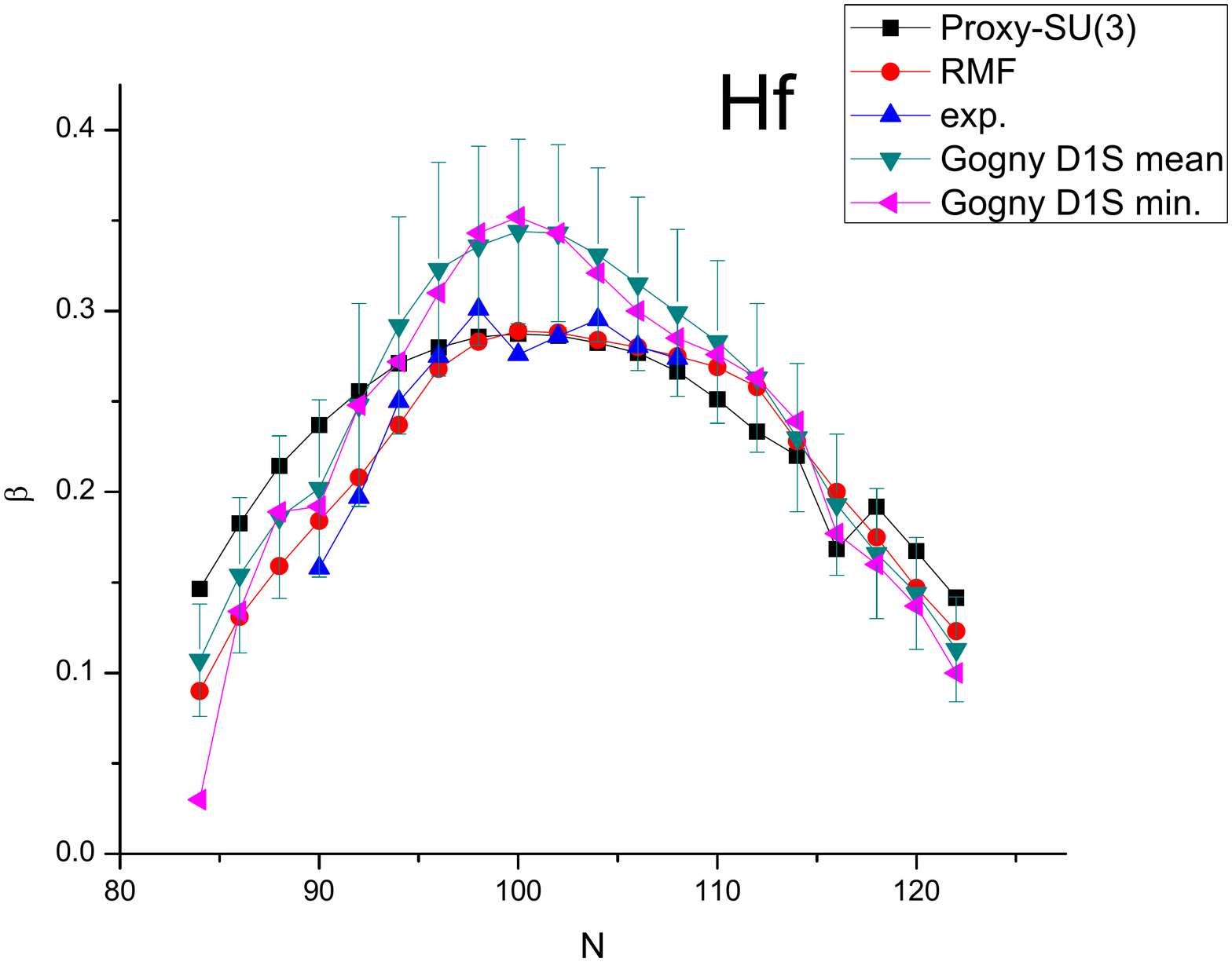}\hspace{5mm}
\includegraphics[width=75mm]{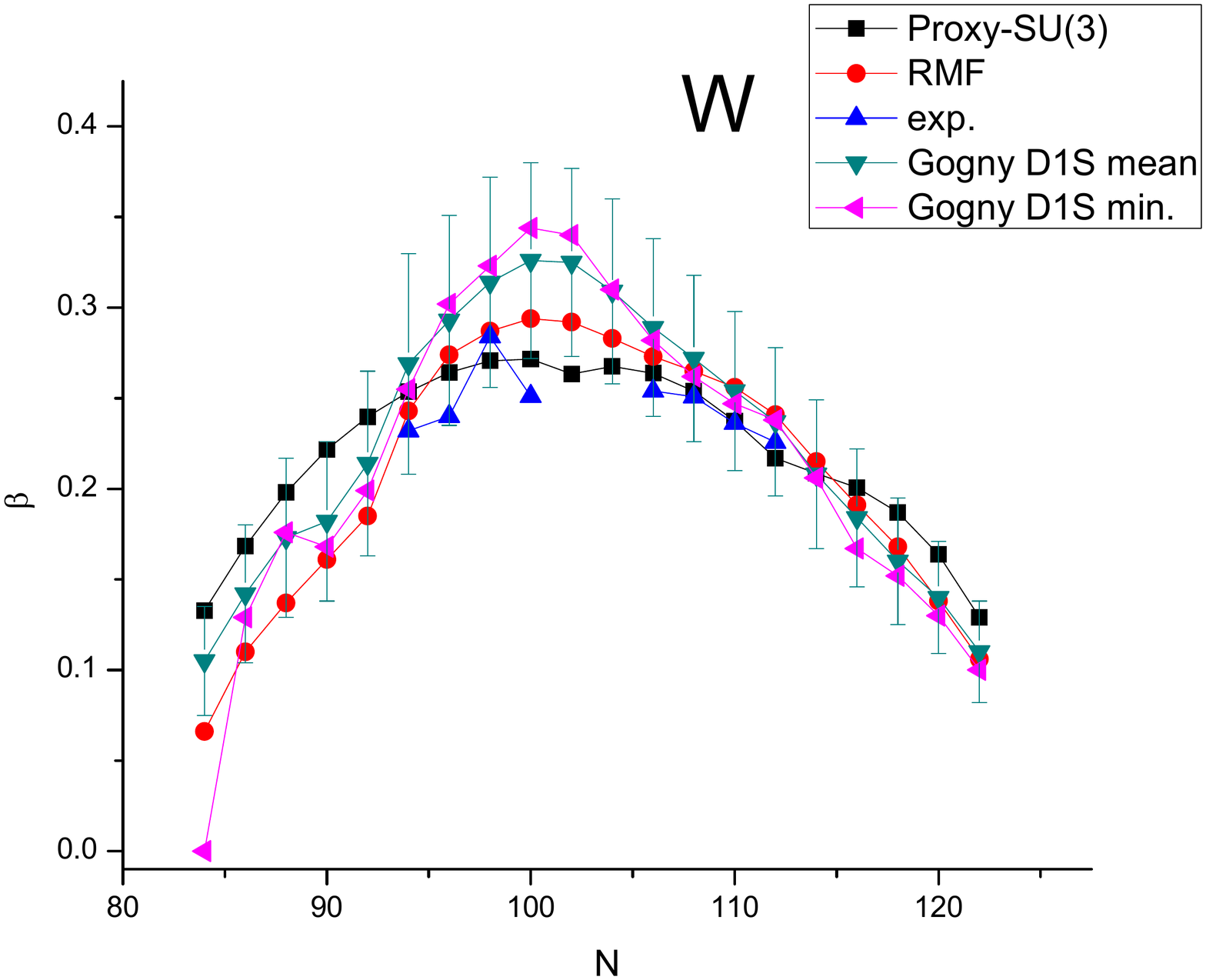}\hspace{5mm}}
{\includegraphics[width=75mm]{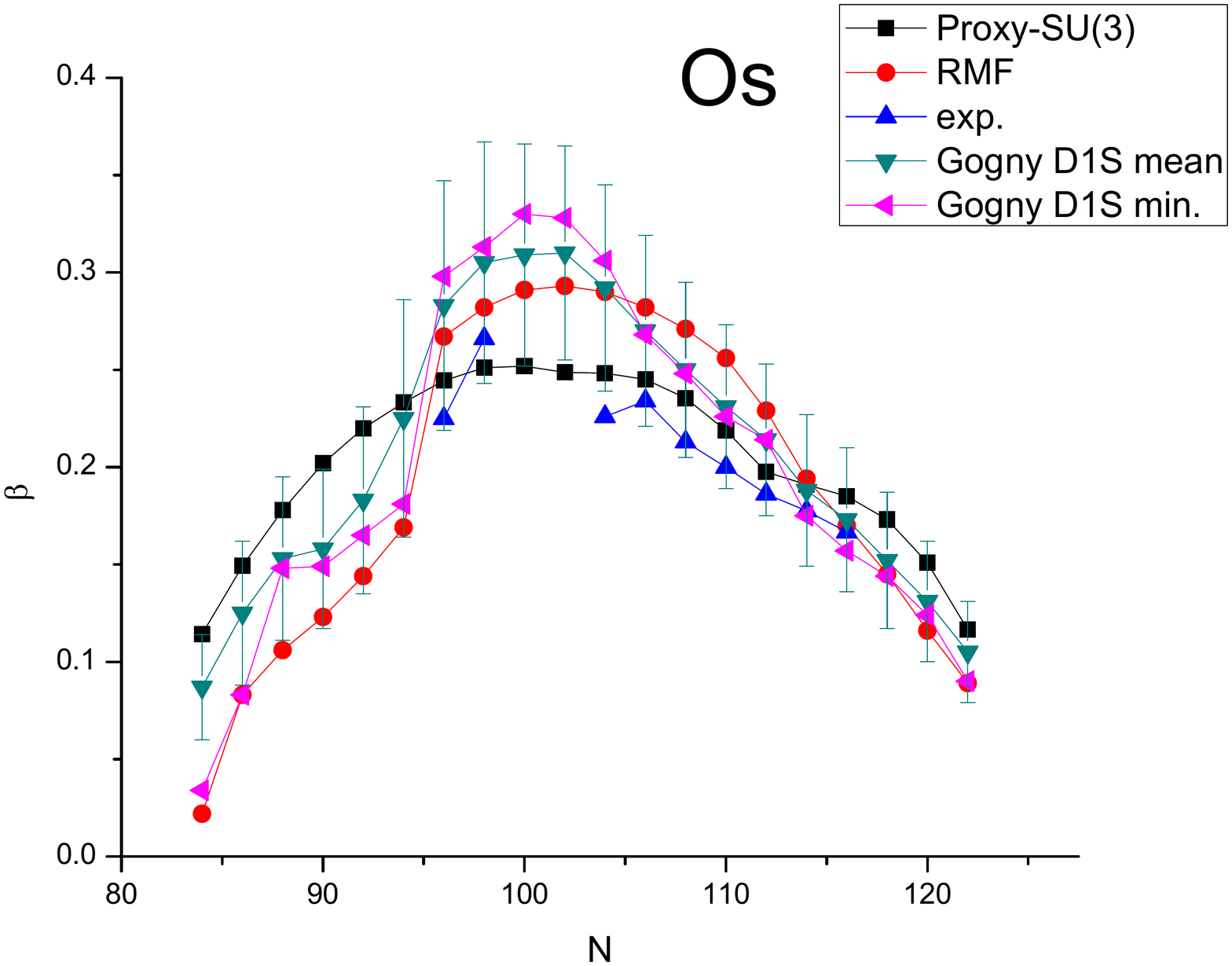}\hspace{5mm}
\includegraphics[width=75mm]{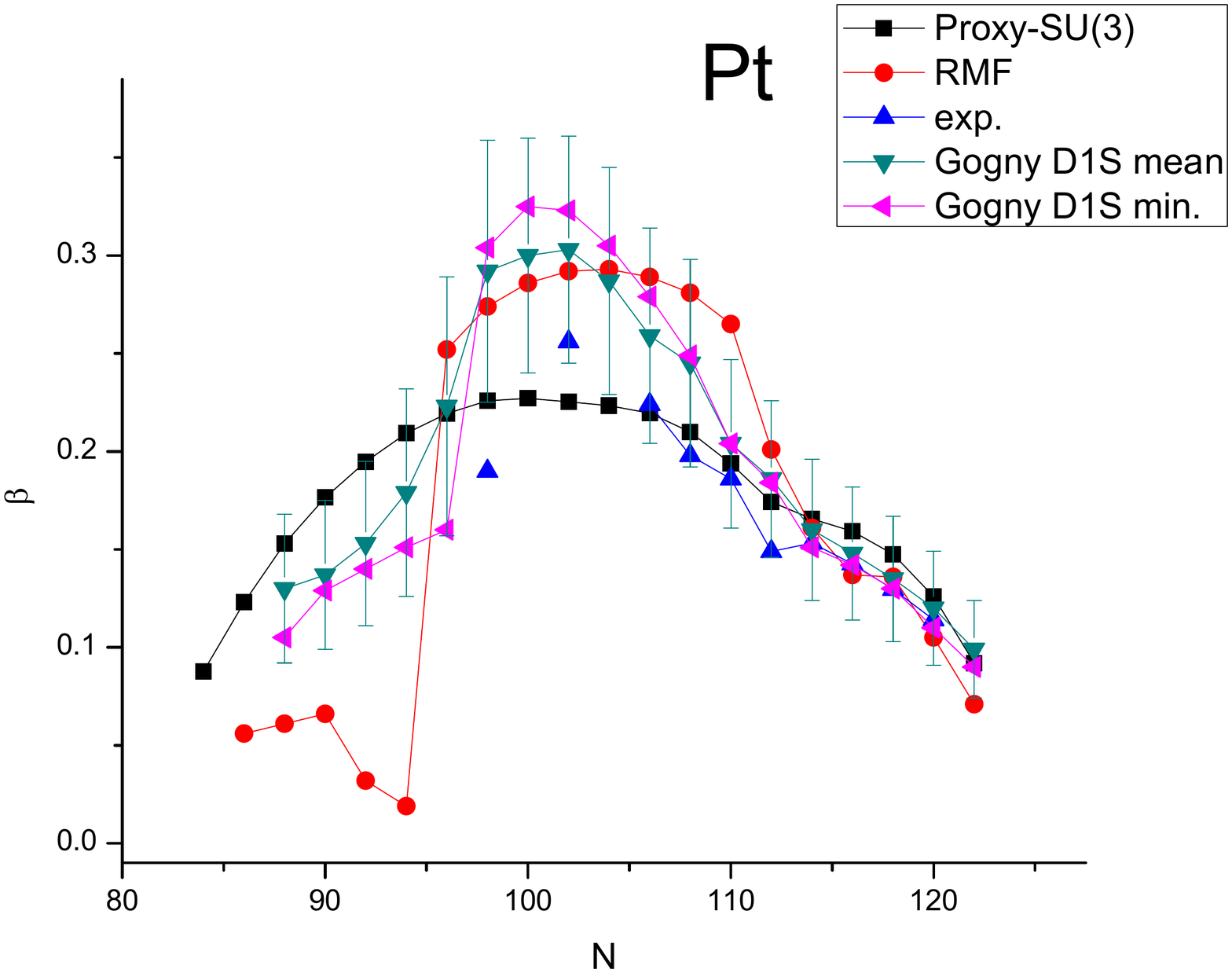}\hspace{5mm}}

\caption{The parameter-free \cite{proxy2} values of the collective variable $\beta$, obtained through the proxy-SU(3) symmetry from Eq. (\ref{b2}), are compared with theoretical results provided by the D1S-Gogny interaction (D1S-Gogny) \cite{Gogny} and by relativistic mean field theory (RMF) \cite{LalazissisADNDT71}, as well as with empirical values (exp.) \cite{Raman}, for the even-even Gd-Pt isotopes. Adapted from Ref. \cite{36J}. See subsection VIII.B for further discussion.} 

\end{figure*}


\begin{figure*}[htb]\label{Fig6}

{\includegraphics[width=75mm]{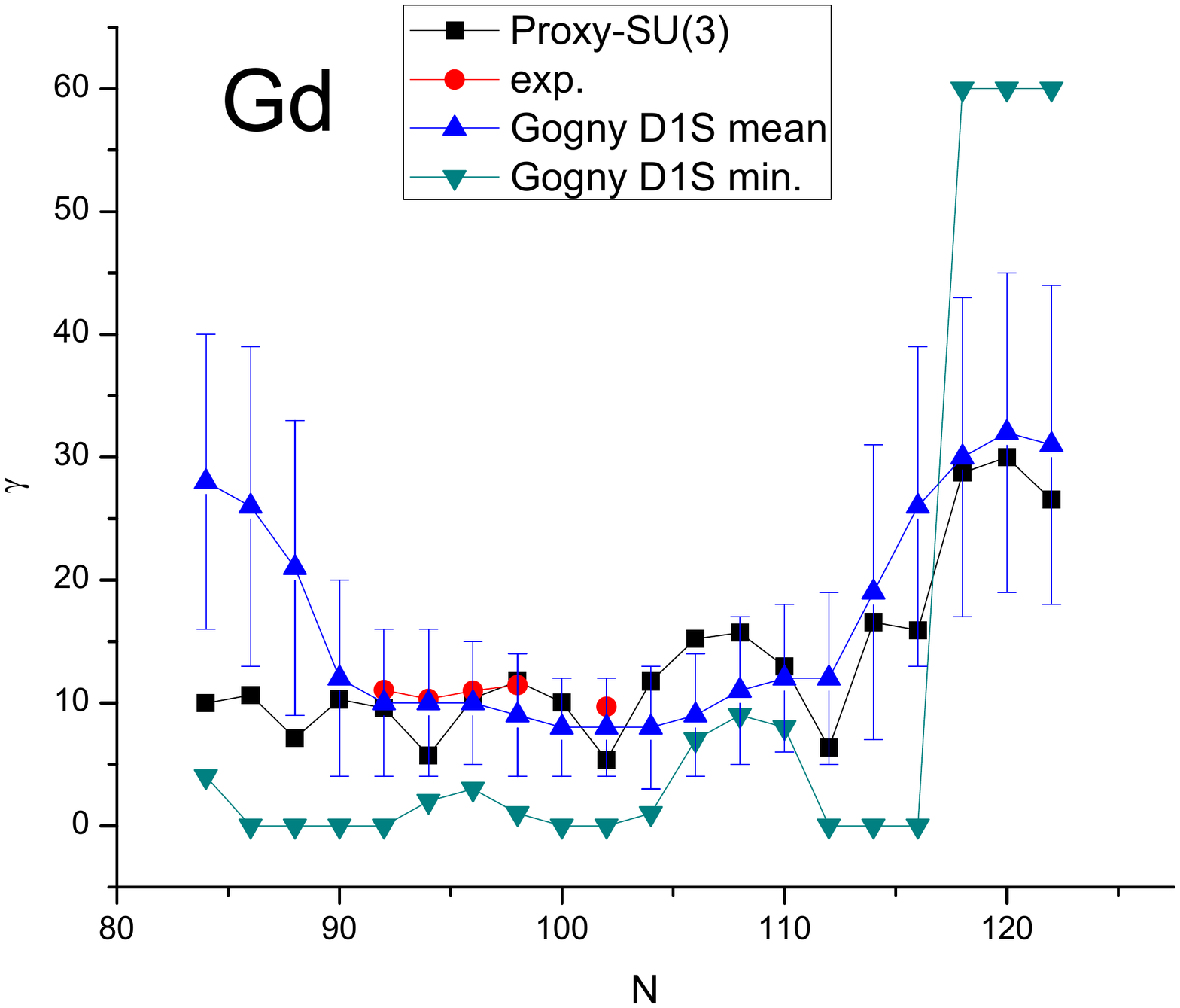}\hspace{5mm}
\includegraphics[width=75mm]{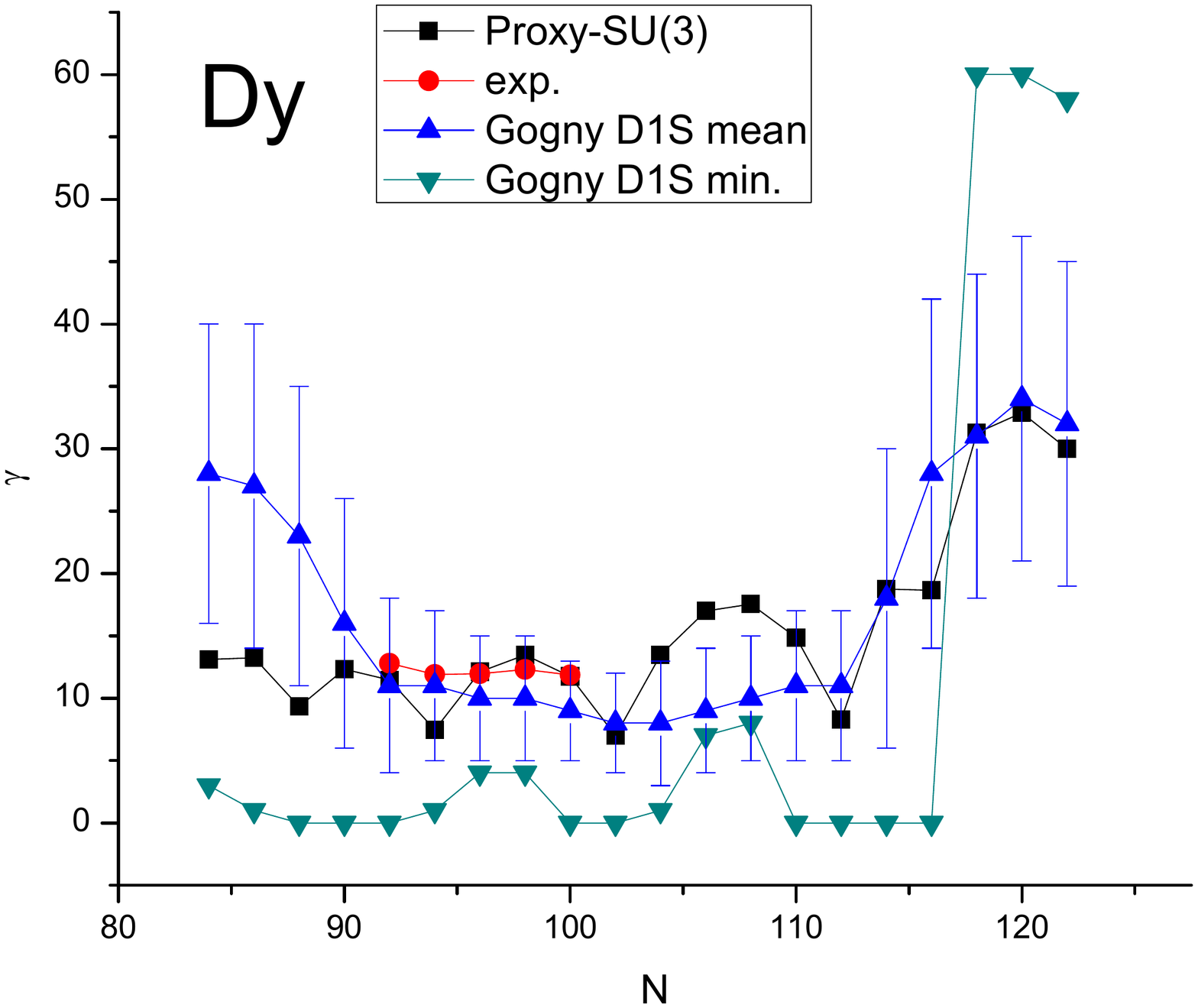}\hspace{5mm}}
{\includegraphics[width=75mm]{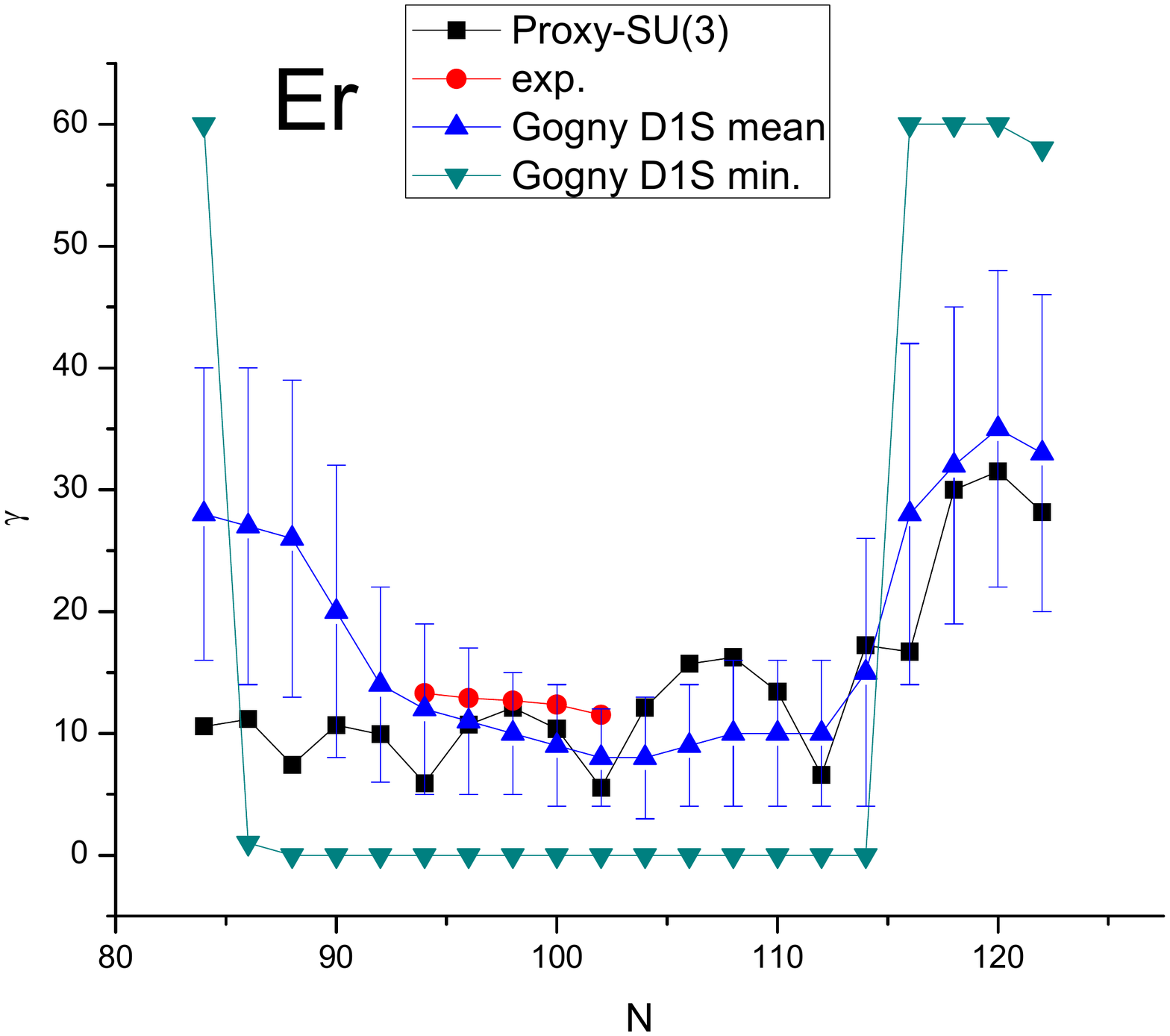}\hspace{5mm}
 \includegraphics[width=75mm]{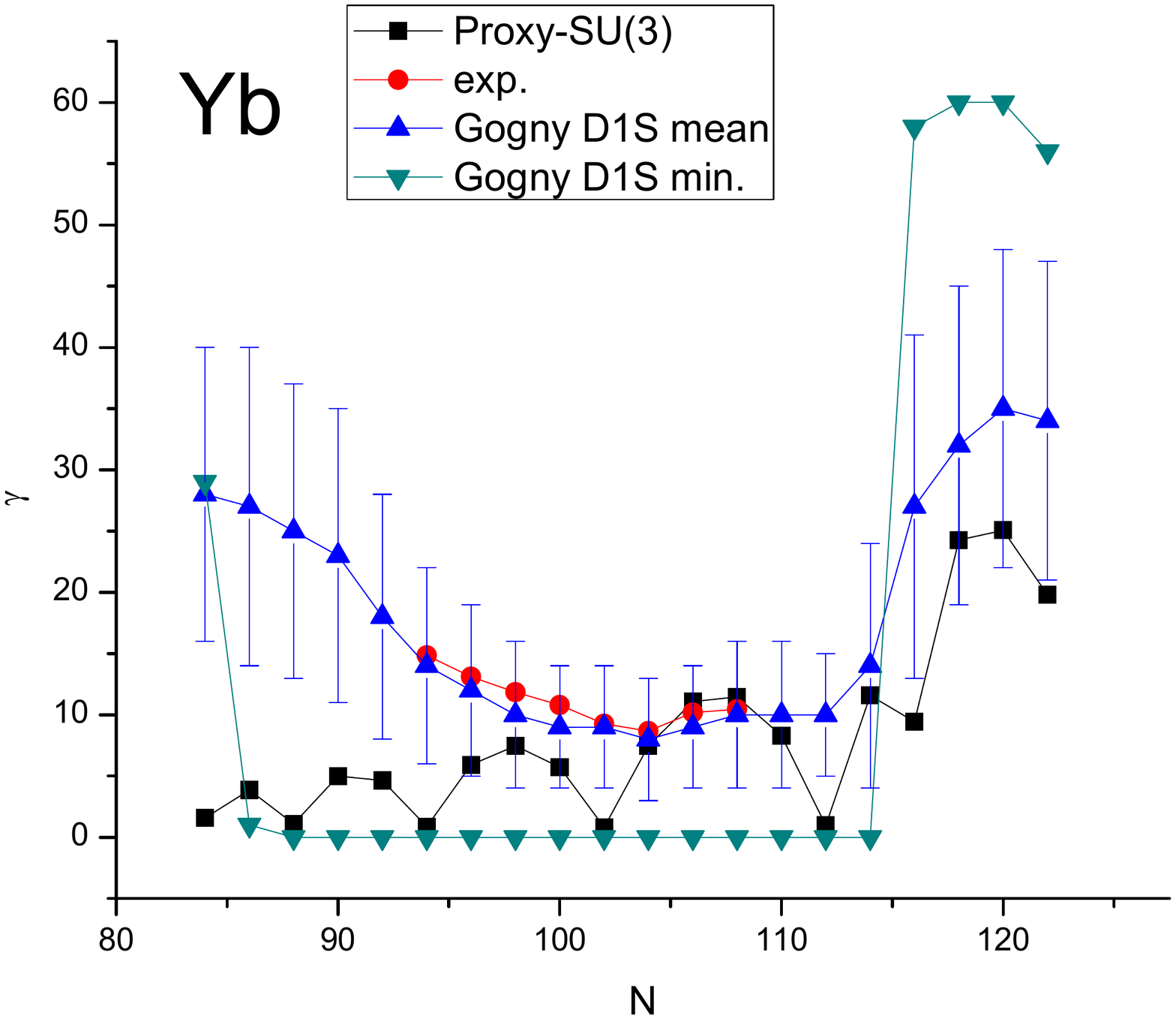}\hspace{5mm}}
{\includegraphics[width=75mm]{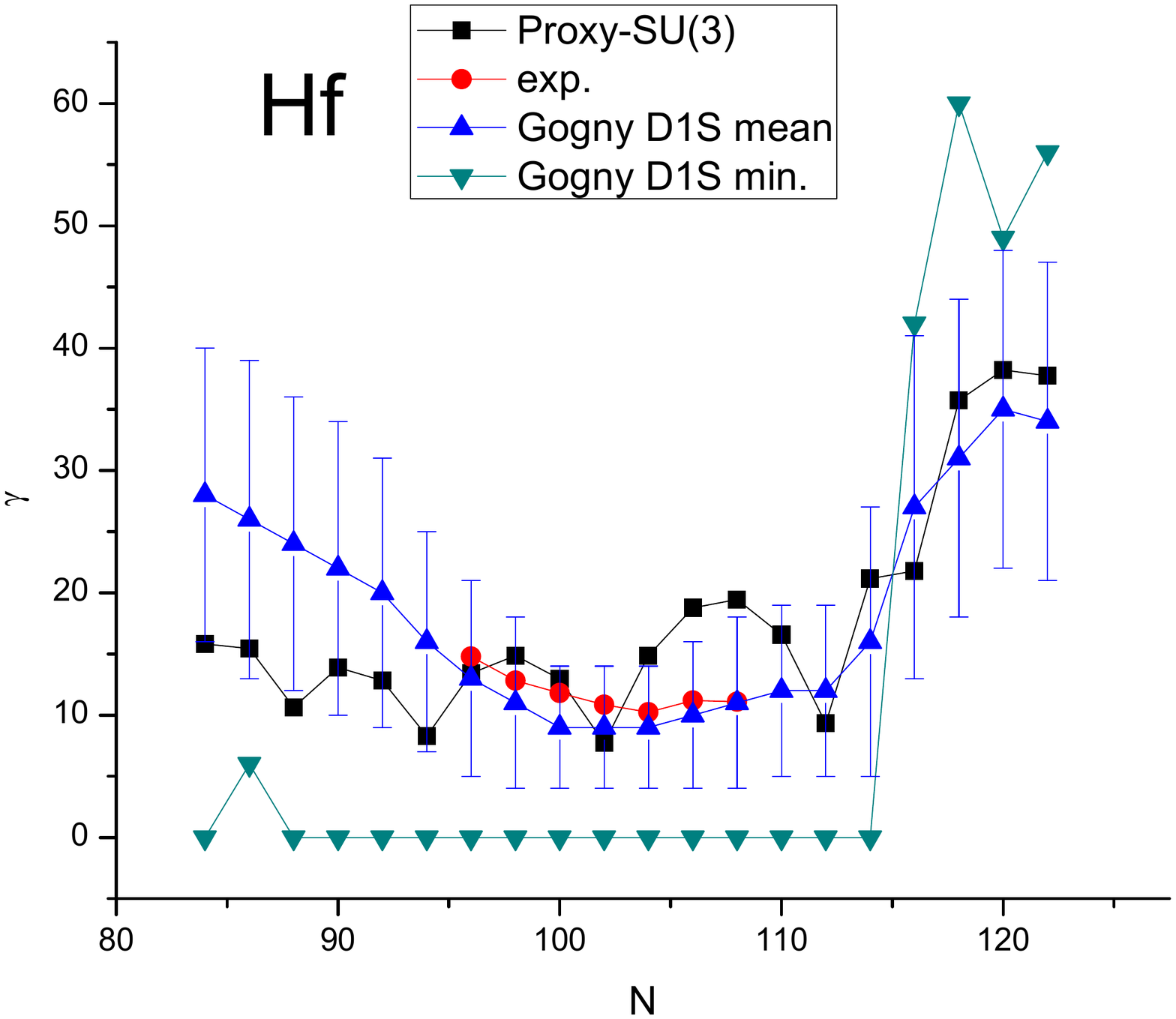}\hspace{5mm}
\includegraphics[width=75mm]{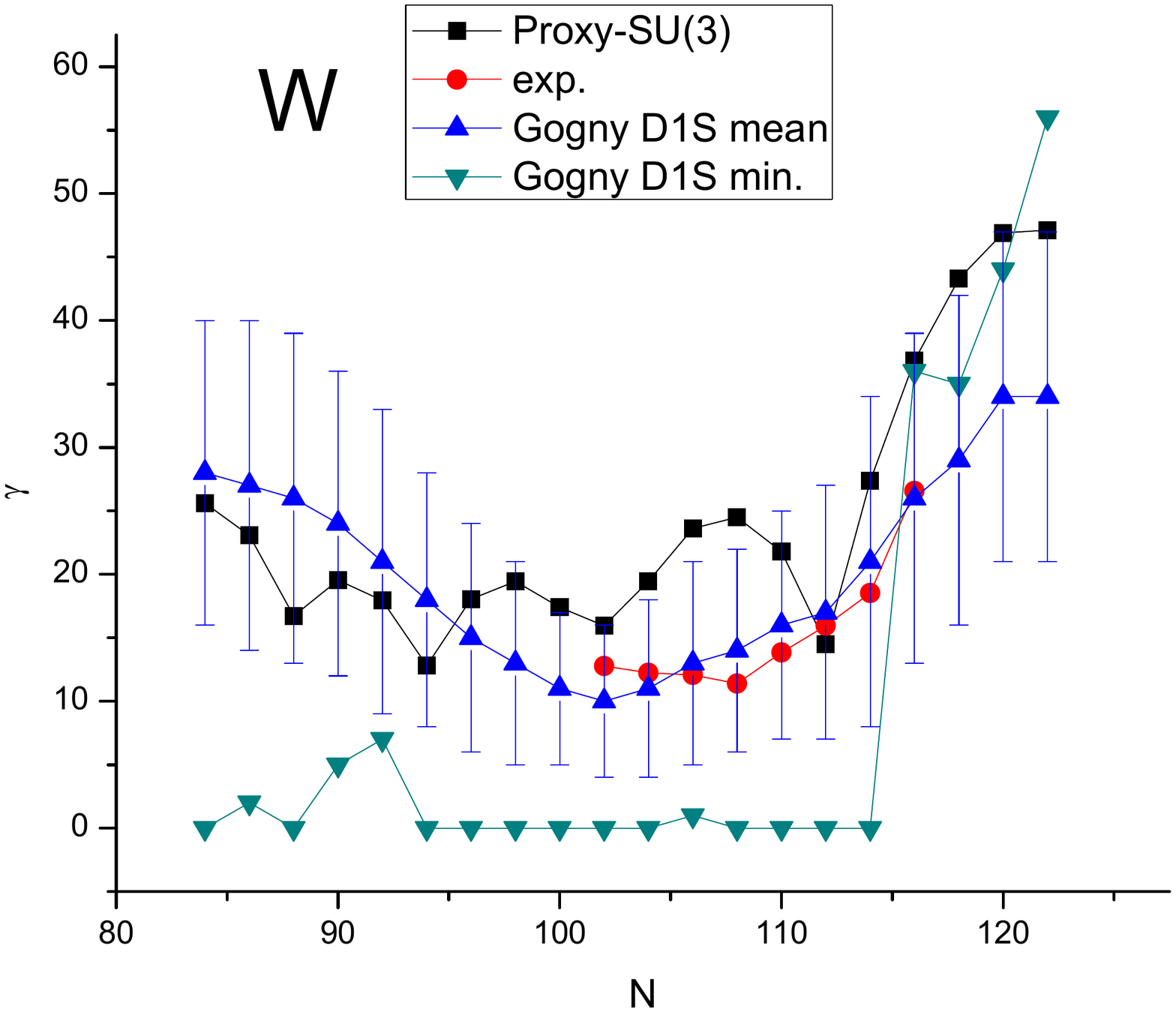}\hspace{5mm}}
{\includegraphics[width=75mm]{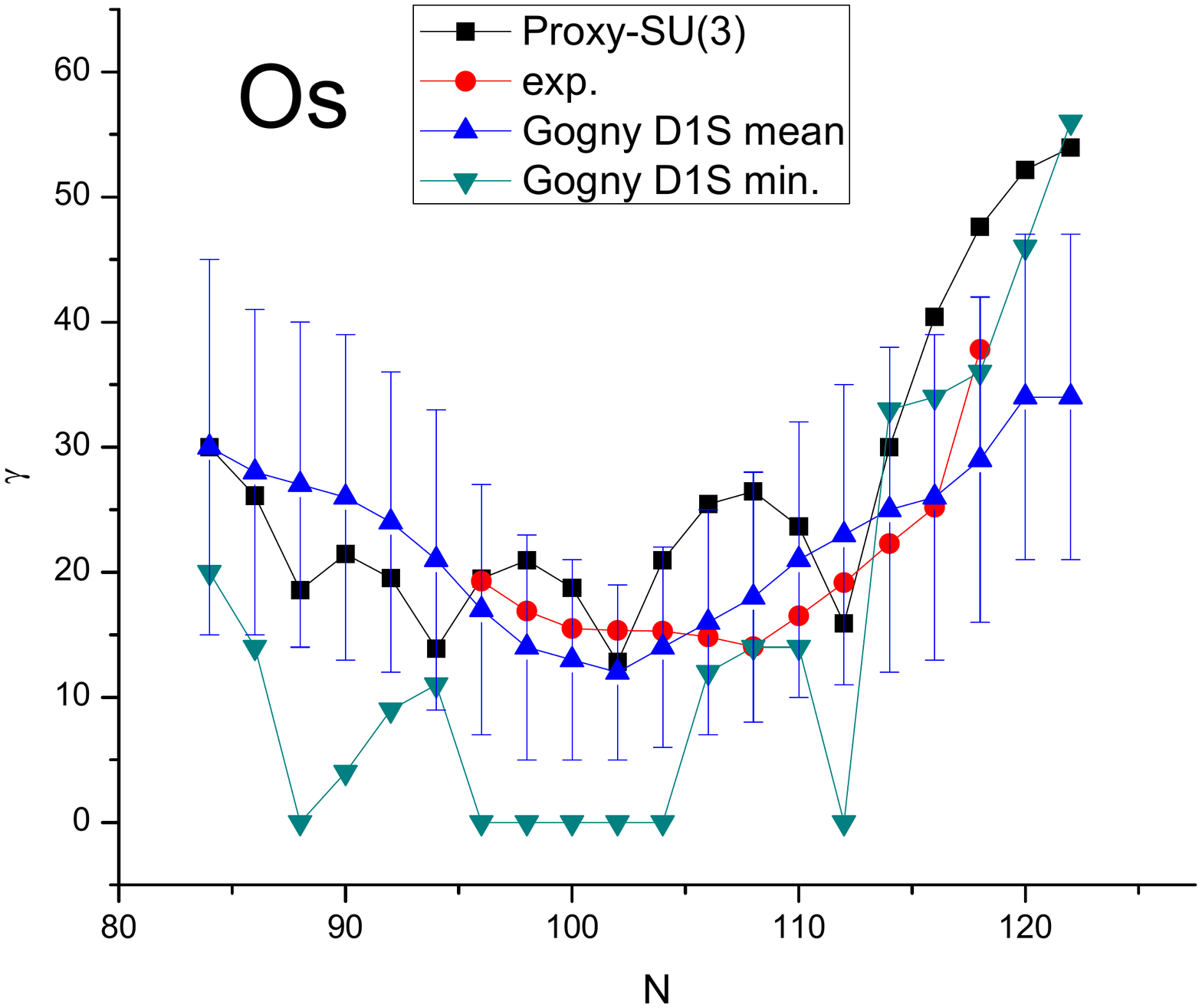}\hspace{5mm}
\includegraphics[width=75mm]{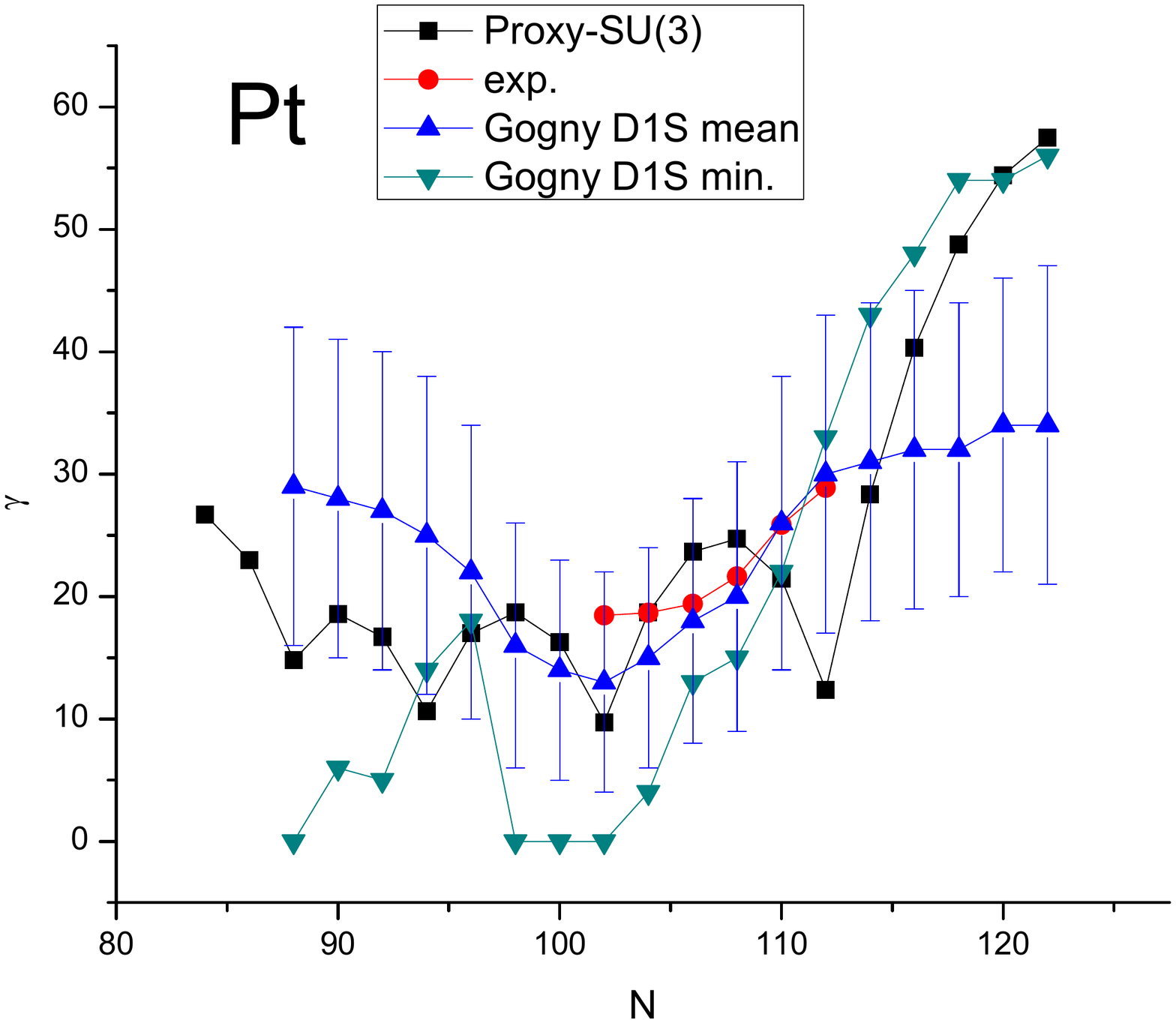}\hspace{5mm}}

\caption{Same as Fig. 4, but for the collective variable $\gamma$, derived from Eq. (\ref{g1}).  Adapted from Ref. \cite{36J}. See subsection VIII.B for further discussion.}

\end{figure*}

\begin{figure}[htb]\label{Fig4}

\includegraphics[width=75mm]{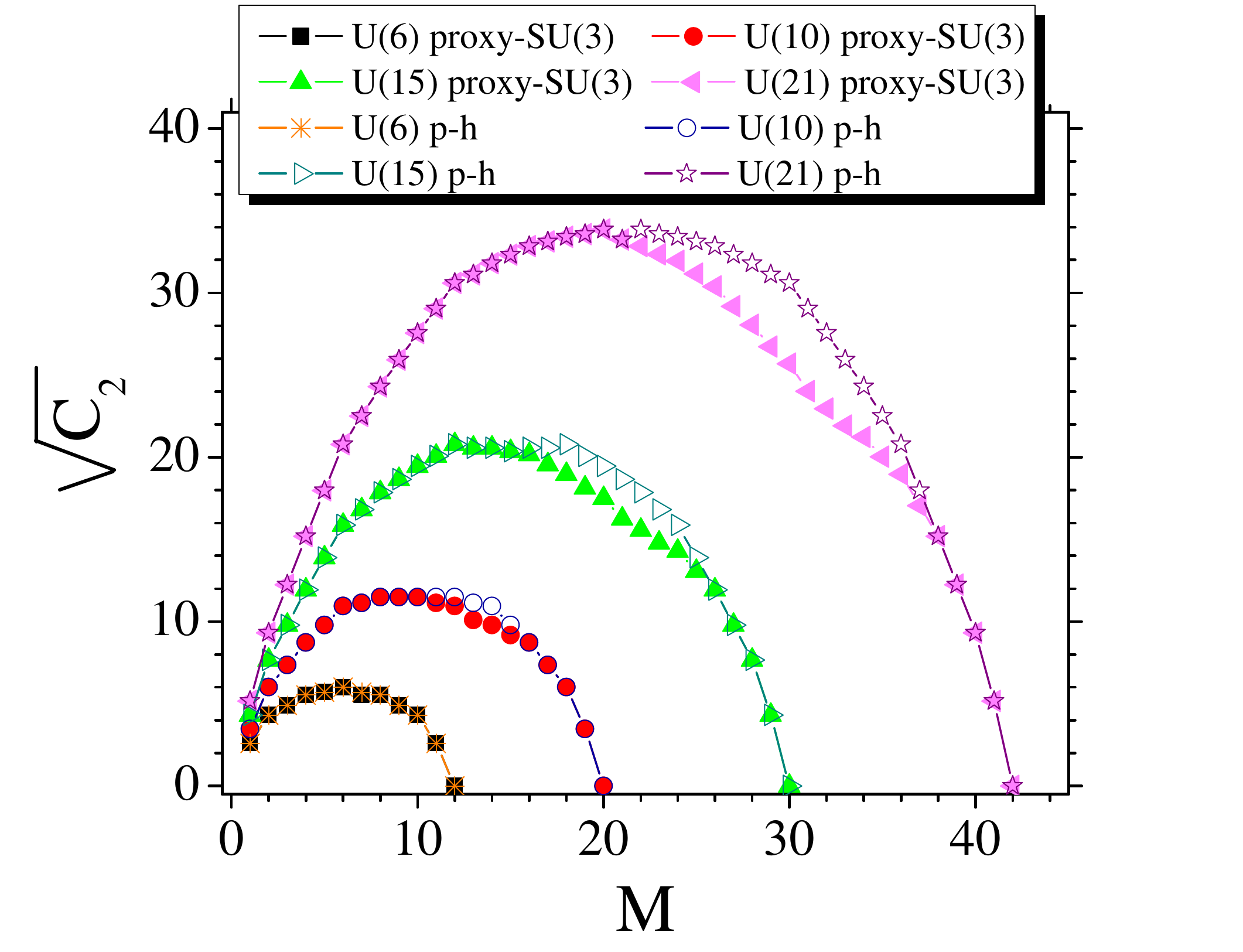}

\caption{The square root of the second order Casimir operator of SU(3), obtained from 
Eq.~(\ref{Casimir}), is shown as a function of the nucleon number M, for different shells. Results obtained by using the proxy-SU(3) irreps (columns hw in Table 6) are labeled by ``proxy-SU(3)'', while those corresponding to the particle-hole symmetry assumption (columns C in Table 6) are labeled by ``p-h''. Adapted from Ref. \cite{proxy2}. See subsection VIII.B  for further discussion.} 
\end{figure}


\begin{figure}[htb]\label{Fig7}

\includegraphics[width=75mm]{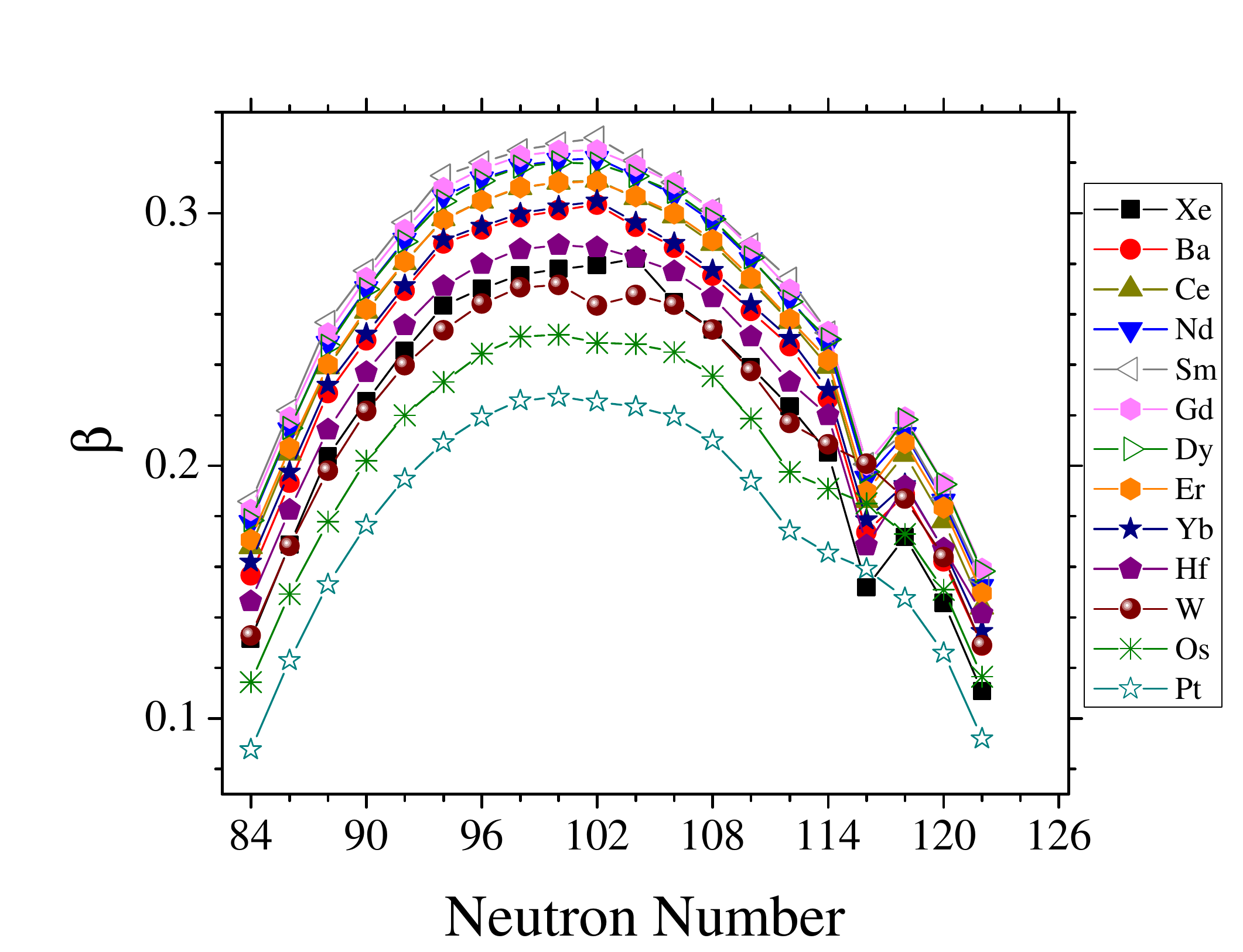}

\caption{Parameter-free \cite{proxy2} proxy-SU(3) predictions for the collective variable  $\beta$,  obtained from Eq. (\ref{b2}). Adapted from Ref. \cite{proxy2}. See Section VIII.C for further discussion.} 

\end{figure}


\begin{figure*}\label{Fig8}
\begin{center}

{\includegraphics[width=75mm]{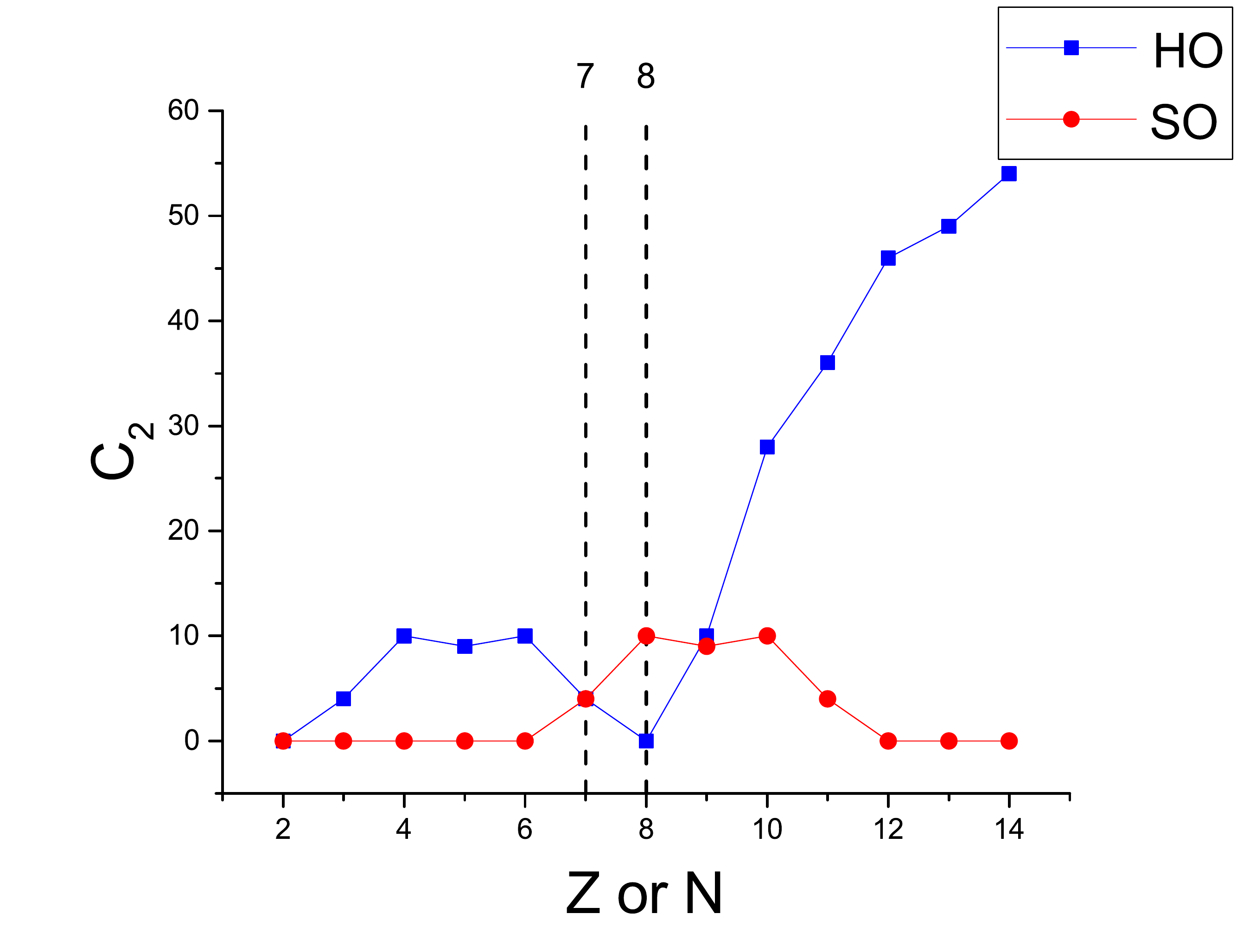}\hspace{5mm}
\includegraphics[width=75mm]{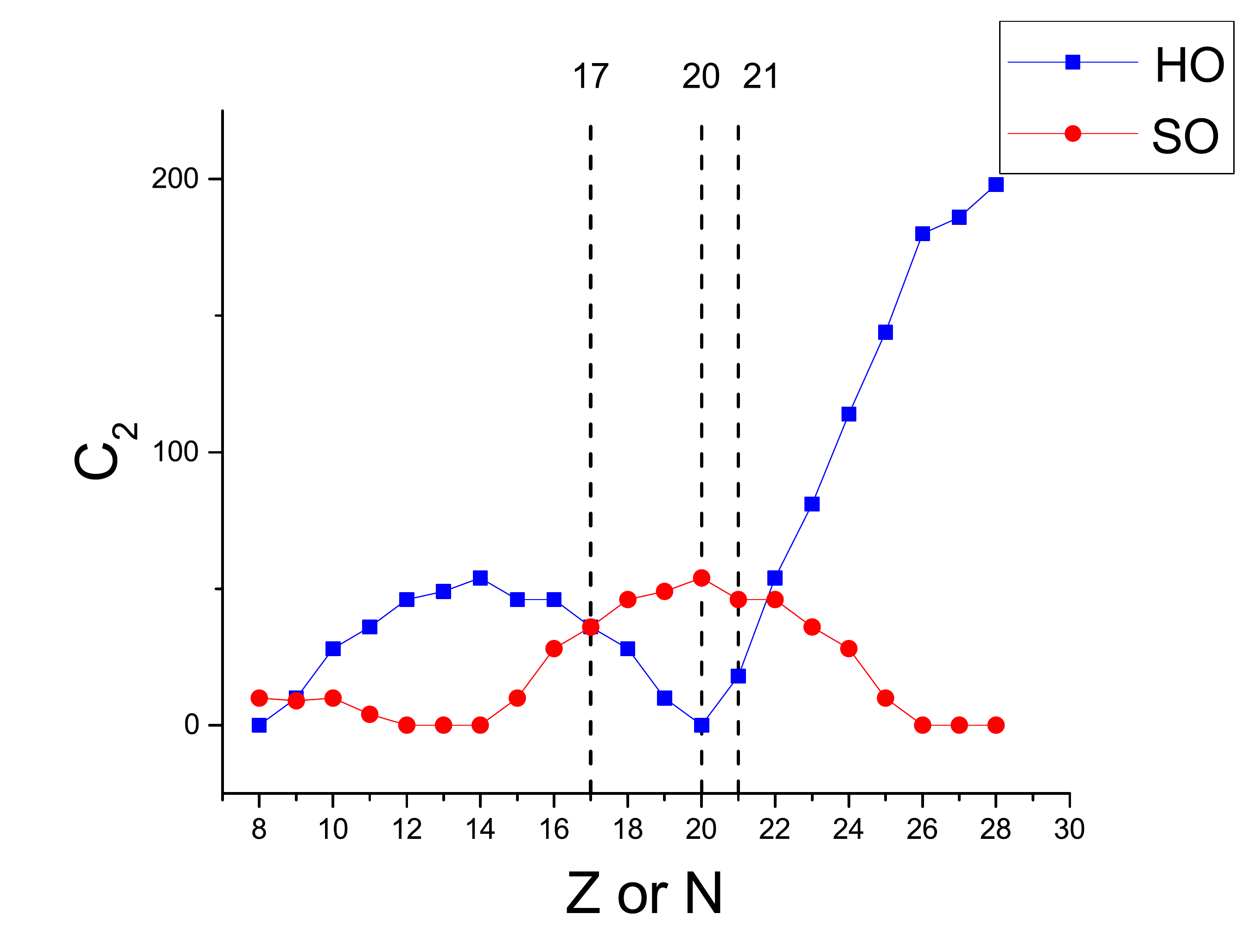}\hspace{5mm}}
{\includegraphics[width=75mm]{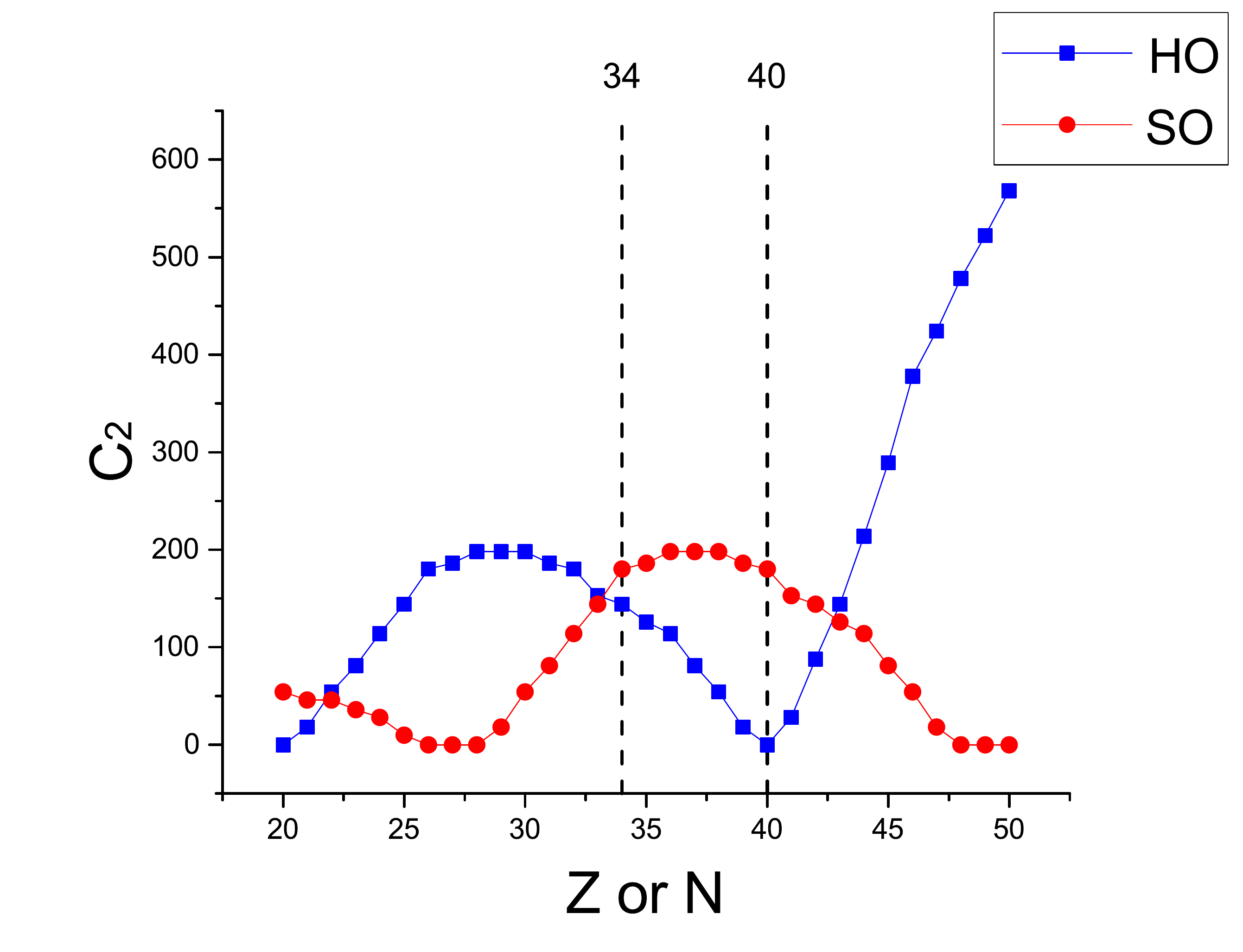}\hspace{5mm}
 \includegraphics[width=75mm]{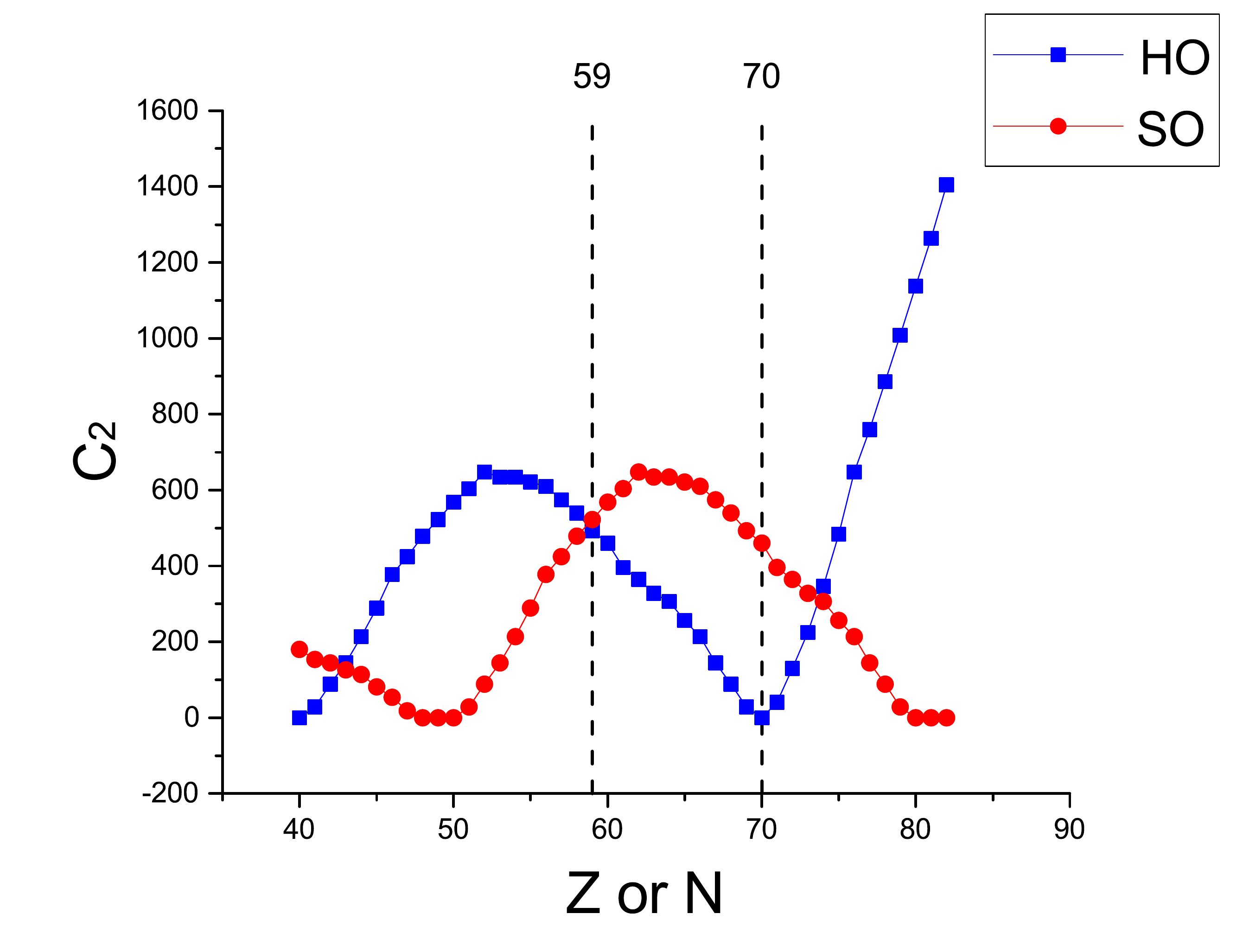}\hspace{5mm}}
{\includegraphics[width=75mm]{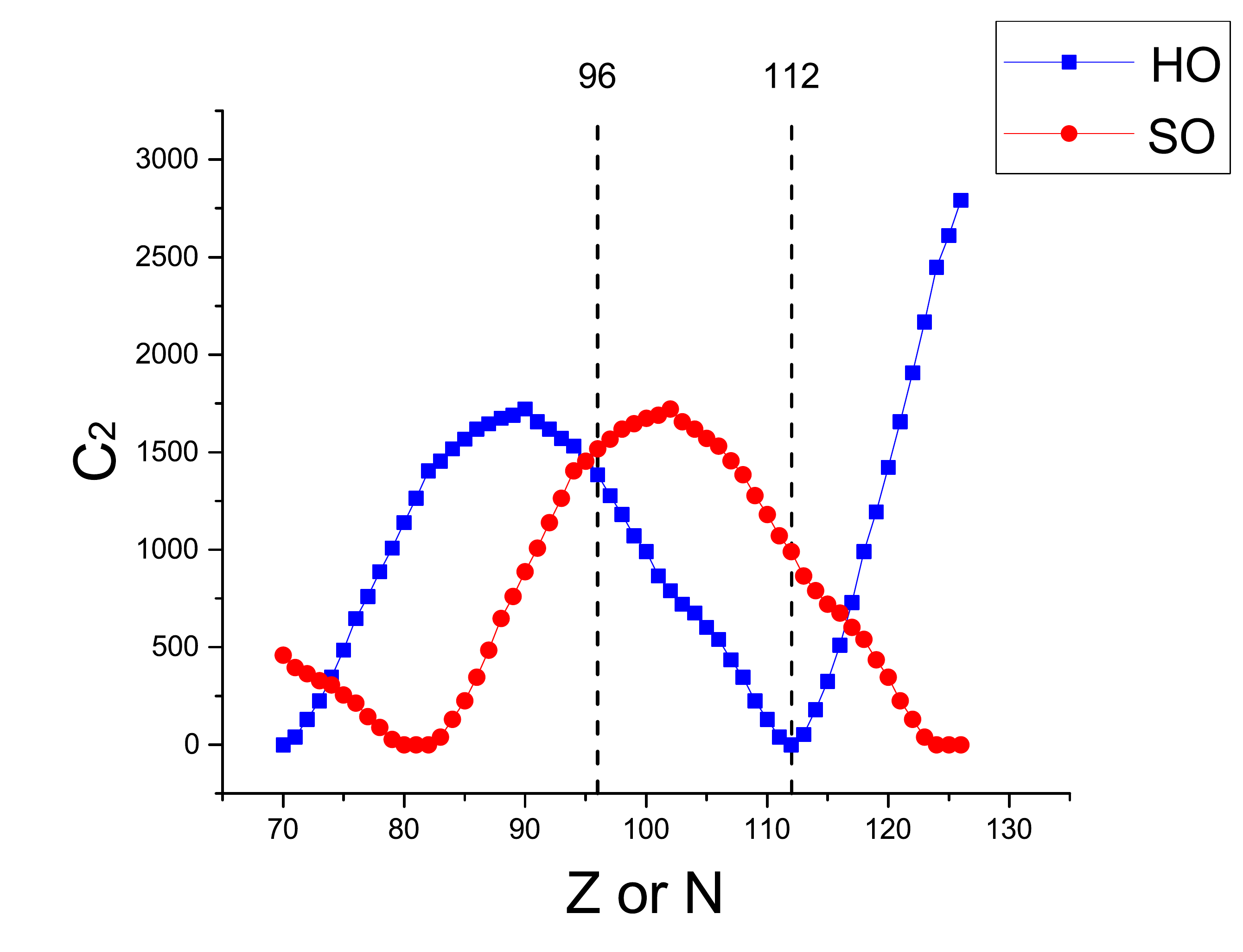}\hspace{5mm}
\includegraphics[width=75mm]{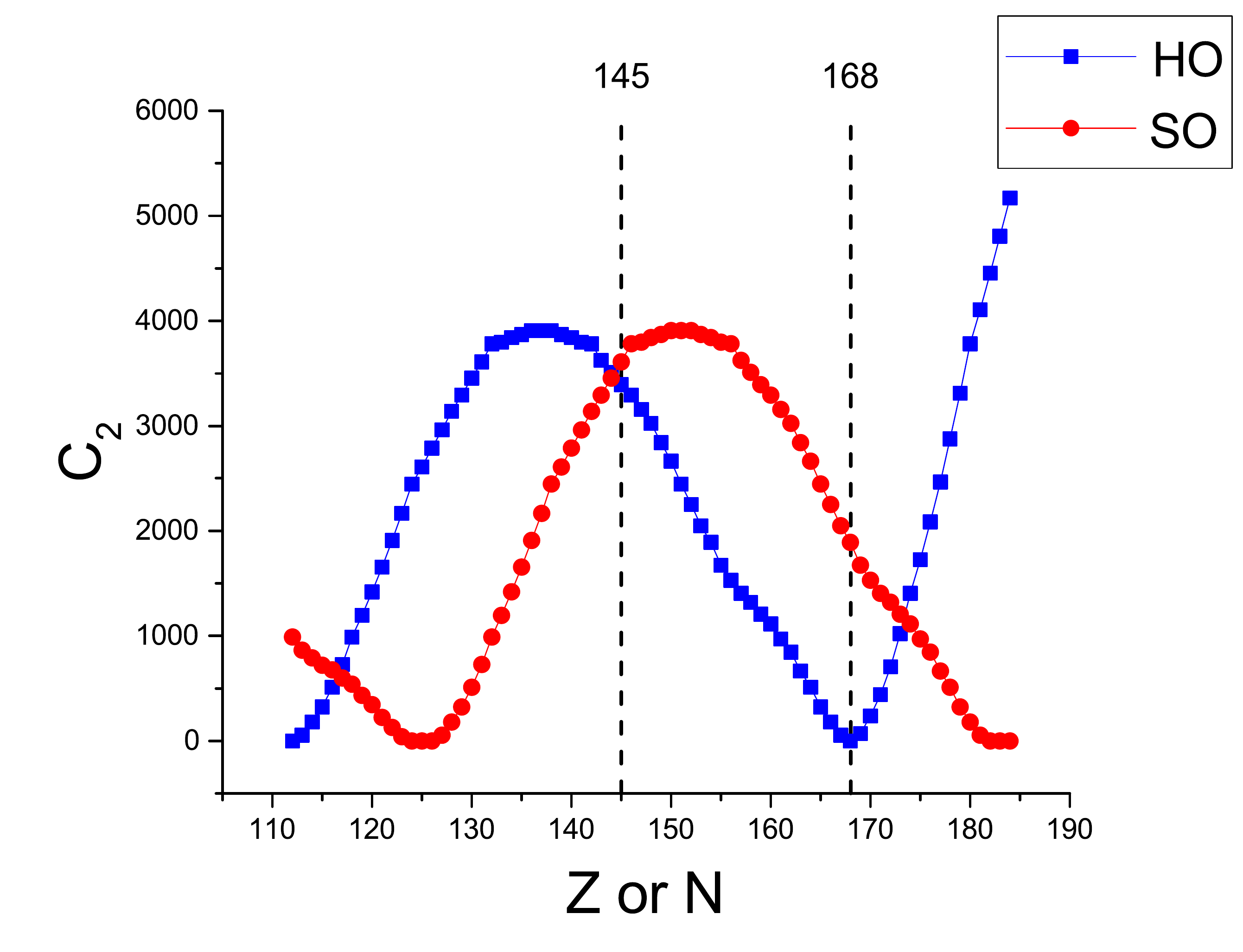}\hspace{5mm}}

\caption{ Eigenvalues of the second order Casimir operator of SU(3) are plotted as a function of the proton ($Z$) or neutron number ($N$). The dual shell mechanism predicts that islands of shape coexistence can occur only when the condition  $C_{2,SO}\ge C_{2,HO}$ is fulfilled. This happens within the proton or neutron number intervals  7-8, 17-20, 34-40, 59-70, 96-112, 145-168. Adapted from Ref. \cite{EPJASC}. See subsection IX.B for further discussion. }\label{isl}
\end{center}
\end{figure*}


\begin{turnpage}
\begin{figure*}[htb]\label{Fig9}
    \centering
    \includegraphics[width=225mm]{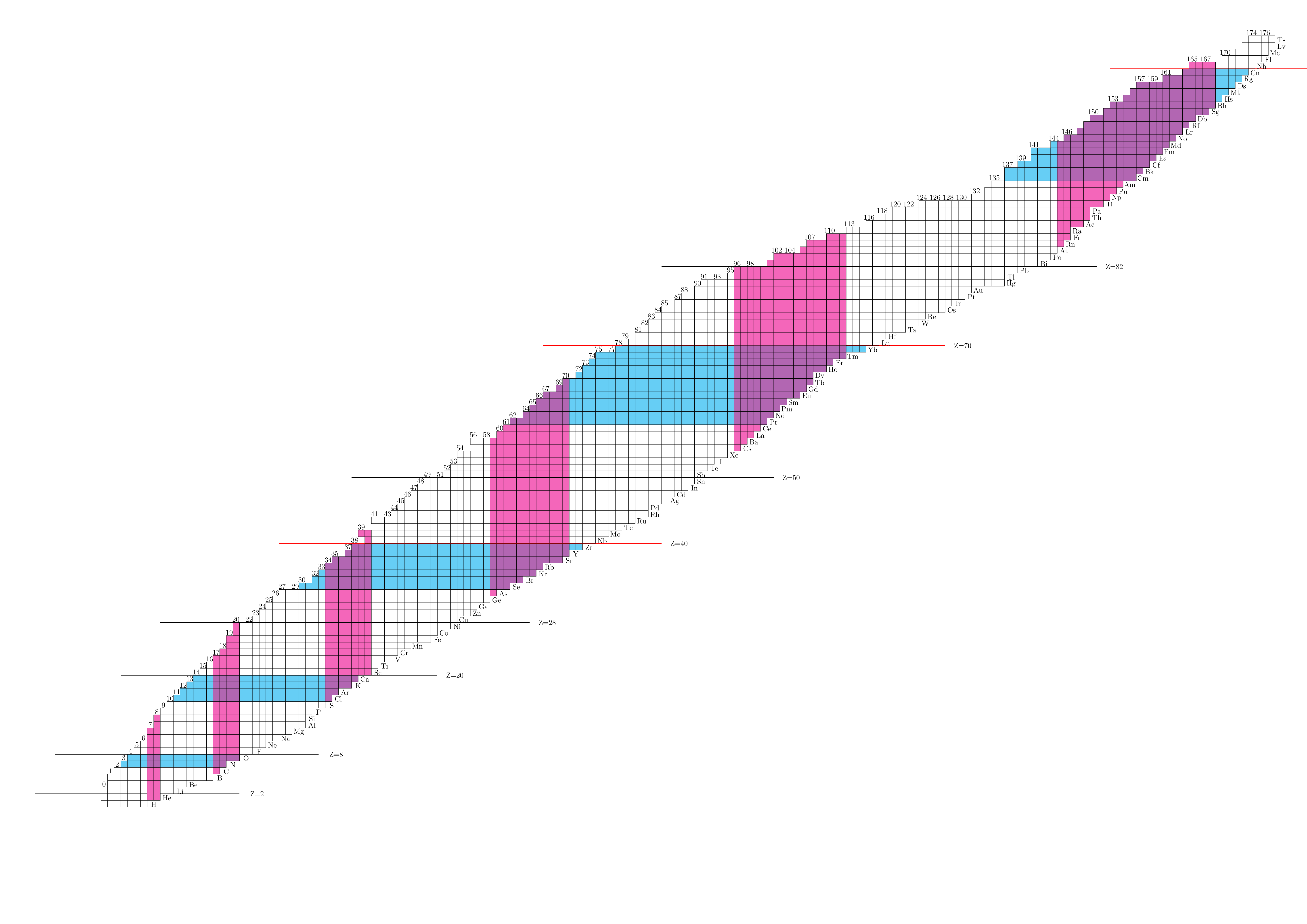}
    \caption{The colored regions in this map possess proton or neutron number within the intervals  7-8, 17-20, 34-40, 59-70, 96-112, 145-168. The map indicates the regions in which shape coexistence can appear, according to the proposed dual shell mechanism. The horizontal stripes correspond to the neutron induced shape coexistence, while the vertical stripes correspond to the proton induced shape coexistence.  Adapted from Ref. \cite{EPJASC}. See section IX for further discussion. }
    \label{map}
\end{figure*}
\end{turnpage}

\end{document}